\documentclass[aps,amsmath,amssymb,twocolumn,pra,superscriptaddress,nofootinbib]{revtex4-2}

\usepackage{graphicx,tikz}
\usepackage{dcolumn}
\usepackage{bm}
\usepackage{epsfig,hyperref,amsthm}
\usepackage{mathtools}
\usepackage{color,fancybox}
\usepackage{colordvi}
\usepackage{relsize}
\usepackage{accents}
\usepackage{wasysym} 
\usepackage{enumitem}
\usepackage{mathtools,slashed}
\usepackage{times}
\usepackage{newtxmath}

\hypersetup{
	colorlinks=true,
	linkcolor=CitingColor,
	citecolor=CitingColor,
	urlcolor=CitingColor
}

\DeclareMathAlphabet\mathbfcal{OMS}{cmsy}{b}{n}
\newcommand{\ket}[1]{\ensuremath{|#1\rangle}}
\newcommand{\bra}[1]{\ensuremath{\langle #1|}}

\newcommand{\proj}[1]{\ket{#1}\bra{#1}}

\newcommand{\norm}[1]{\left\|#1\right\|}

\newcommand{\id}{\mathbb{I}}

\newtheorem{theorem}{Theorem}
\newtheorem{corollary}{Corollary}
\newtheorem{lemma}{Lemma}
\newtheorem{proposition}{Proposition}

\newtheorem{question}{Question}
\newtheorem{fact}{Fact}

\newtheorem{definition}{Definition}

\definecolor{nred}{rgb}{0.9,0.1,0.1}
\definecolor{nblack}{rgb}{0,0,0}
\definecolor{nblue}{rgb}{0.2,0.2,0.8}
\definecolor{ngreen}{rgb}{0.2,0.6,0.2}
\definecolor{ublue}{rgb}{0,0,0.5}

\definecolor{pur}{rgb}{0.75,0,0.75}
\definecolor{nngrn}{rgb}{0,0.5,0.5}
\definecolor{CitingColor}{rgb}{0,0.3,1}

\newcommand{\CY}[1]{{\color{black}#1}}
\newcommand{\CYnew}[1]{{\color{black}#1}}

\begin{document}
\title{Dynamical resource theory of incompatibility preservability}

\author{Chung-Yun Hsieh}
\affiliation{H. H. Wills Physics Laboratory, University of Bristol, Tyndall Avenue, Bristol, BS8 1TL, UK}

\author{Benjamin Stratton}
\affiliation{Quantum Engineering Centre for Doctoral Training, H. H. Wills Physics Laboratory and Department of Electrical \& Electronic Engineering, University of Bristol, Bristol BS8 1FD, UK}
\affiliation{H. H. Wills Physics Laboratory, University of Bristol, Tyndall Avenue, Bristol, BS8 1TL, UK}

\author{Chao-Hsien Wu}
\affiliation{Department of Physics, National Taiwan Normal University, Taipei 11677, Taiwan}

\author{Huan-Yu Ku}
\email{huan.yu@ntnu.edu.tw}
\affiliation{Department of Physics, National Taiwan Normal University, Taipei 11677, Taiwan}

\date{\today}

\begin{abstract}
The uncertainty principle is one of quantum theory's most foundational features. 
It underpins a quantum phenomenon called {\em measurement incompatibility}---two physical observables of a single quantum system may not always be measured simultaneously.
Apart from being fundamentally important, measurement incompatibility is also a powerful {\em resource} in the broad quantum science and technologies, with wide applications to cryptography, communication, random number generation, and device-independent tasks. Since every physical system is unavoidably subject to noise, an important, yet still open, question is how to characterise the ability of noisy quantum dynamics to {\em preserve} measurement incompatibility.
This work fills this gap by providing the first resource theory of this ability, termed {\em incompatibility preservability}.
We quantify incompatibility preservability by a robustness measure.
Then, we introduce an operational task, {\em entanglement-assisted filter game}, to completely characterise both the robustness measure and the conversion of incompatibility preservability.
Our results provide a general framework to describe how noisy dynamics affect the uncertainty principle's signature.
\end{abstract}

\maketitle

\section{Introduction}
The uncertainty principle is one of the most profound aspects of quantum theory~\cite{Busch2014RMP}.
It describes an intriguing quantum phenomenon termed {\em measurement incompatibility}~\cite{Otfried2023RMP}---two physical observables of a single quantum system may not always be measured simultaneously (and, thus, are {\em incompatible}).
For instance, spin measurements in two orthogonal directions are incompatible, as measuring one will unavoidably prevent us from knowing the other.
It turns out that measurement incompatibility is a powerful {\em quantum resource}---by using it, one can achieve things {\em unachievable} in its absence~\cite{Otfried2023RMP}.
For instance, in quantum cryptography, the well-known BB84 protocol~\cite{BB84} needs a pair of incompatible measurements.
Also, measurement incompatibility is necessary to demonstrate Bell nonlocality~\cite{Brunner2014RMP,Quintino2014PRL,Cavalcanti2016PRA}, quantum steering~\cite{UolaRMP2020,Cavalcanti2016,Quintino2014PRL,Uola2014PRL,Uola2015PRL,Cavalcanti2016PRA,Zhao2020}, and quantum complementarity~\cite{Otfried2023RMP,Hsieh-IP}. 
Nowadays, measurement incompatibility has been recognised as an indispensable resource in, e.g., quantum cryptography~\cite{BB84,Hsieh-IP,Acin2007PRL,Branciard2012PRA,Pironio2009NJP,Acin2006PRL}, quantum communication~\cite{Buscemi2023Quantum,Buscemi2020PRL,Ji2024PRXQ,Skrzypczyk2019PRL}, (semi-)device-independent quantum-informational tasks~\cite{Ku2022NC,ku2023,Hsieh2023,Branciard2012PRA}, random number generation~\cite{Pironio2010Nature,SkrzypczykPRL2018}, and thermodynamics~\cite{Hsieh2024,Hsieh2023IP,KosloffPRE2002,FeldmannPRE2006,LostaglioNJP2017,Majidy2023,YungerHalpern2016NC,Guryanova2016NC,JiPRL2022,BeyerPRL2019,ChanPRA2022}.

Crucially, every practical system is subject to noise. 
An important question is: {\em After noisy dynamics, can we still detect measurement incompatibility?}
A thorough answer to such a question is vital for practical applications.
For instance, studying noisy channels' ability to maintain/transmit information leads to {\em quantum communication} theory (see, e.g., Ref.~\cite{wilde_2017}). 
Understanding how local dynamics maintain global entanglement initiates the resource theory of {\em quantum memory}~\cite{RossetPRX2018,Ku2022PRXQ,Yuan2021npjQI,Vieira2024,Chang2024PRR}. 
Investigating how to preserve general quantum resources via channels gives {\em resource preservability}~\cite{Hsieh2020Quantum,Hsieh2021PRXQ}, which has been successfully applied to study the preservation of informational non-equilibrium~\cite{,Stratton2024PRL} (see also Refs.~\cite{SaxenaPRR2020,Hsieh2020PRR,ChenPRA2021}). 
Still, up to now, a thorough framework for fully quantifying and characterising noisy quantum dynamics' ability to {\em preserve} measurement incompatibility is still missing.

This work fills this gap by providing the first such framework via a dynamical resource theory of {\em incompatibility preservability}. 
Conceptually, our framework starts with defining dynamics that remove a system's incompatibility signature as ``free''. 
Then, we quantify dynamics' ability to preserve incompatibility by checking its distance from being free.

\CYnew{\section{Framework}}

\subsection{Quantum measurements and their incompatibility}
A quantum measurement can be described by a {\em positive operator-valued measure} (POVM)~\cite{QIC-book} $\{E_a\}_a$ satisfying $E_a~\ge~0$ $\forall\,a$ and $\sum_aE_a = \id$.
Physically, it describes the process that, when we measure a system initially in a state $\rho$, we obtain the $a$-th outcome with probability ${\rm tr}(E_a\rho)$.
Note that POVMs only output classical statistics of measurement outcomes.
\CY{Also, in this work, we only consider finite-dimensional systems.}
Now, for a given system, suppose we have several measurements: For each $x$, the set $\{E_{a|x}\}_a$ is a POVM with outcomes indexed by $a$.
The collection \mbox{${\bf E}\coloneqq\{E_{a|x}\}_{a,x}$} including all $a,x$ is called a {\em measurement assemblage}.
This notion describes multiple physical observables in a single system (in this work, ``{\bf E}, {\bf M}'' denote measurement assemblages).
E.g., Pauli X, Z measurements are described by ${\bf E}^{\rm Pauli}\coloneqq\{E_{0|0}^{\rm Pauli}=\proj{0},E_{1|0}^{\rm Pauli}=\proj{1}, E_{0|1}^{\rm Pauli}=\proj{+},E_{1|1}^{\rm Pauli}=\proj{-}\}$, where $\ket{\pm}\coloneqq(\ket{0}\pm\ket{1})/\sqrt{2}$ are eigenstates of Pauli X.

\begin{figure}
\begin{center}
\scalebox{0.8}{\includegraphics{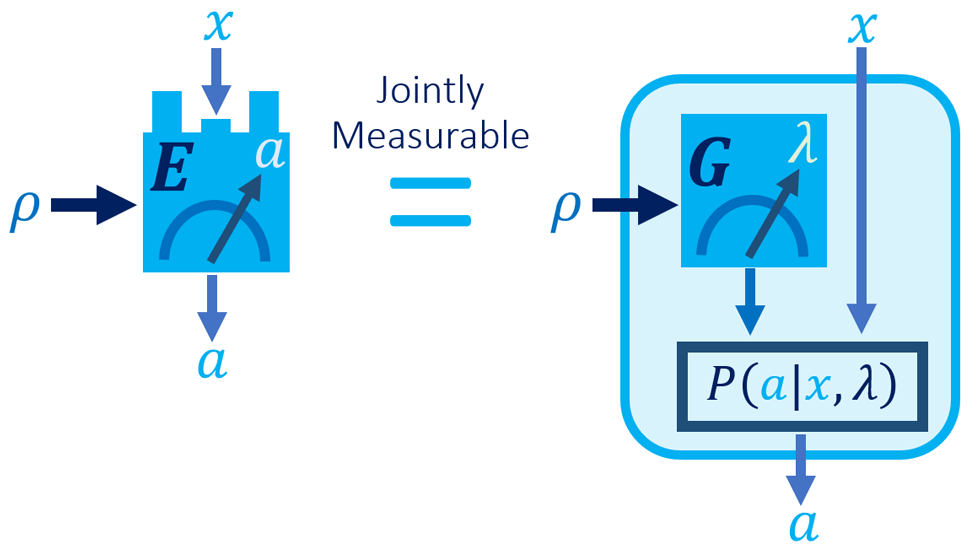}}
\caption{{\bf Jointly-measurable measurements.}
${\bf E}=\{E_{a|x}\}_{a,x}$ is jointly measurable if, for all input state $\rho$, its measurement outcome statistics can be reproduced by a {\em single} POVM ${\bf G}=\{G_\lambda\}_\lambda$ plus classical post-processing $P(a|x,\lambda)$. In the figure, horizontal (vertical) lines are for quantum (classical) information processing.}
\label{Fig:incomp} 
\end{center}
\end{figure}

A foundationally important question for multiple measurements is: {\em Can they be jointly measured?}
Formally, we say a measurement assemblage ${\bf E}$ is {\em jointly-measurable}~\cite{Ali2009FP,Uola2014PRL,Otfried2023RMP} if there exists a {\em single} POVM $\{G_\lambda\}_\lambda$ and some conditional probability distributions $\{P(a|x,\lambda)\}_{a,x,\lambda}$~\footnote{Namely, $P(a|x,\lambda)\ge0$ $\forall\,a,x,\lambda$; $\sum_aP(a|x,\lambda)=1$ $\forall\,x,\lambda$.} such that
\begin{align}\label{Eq:JM}
E_{a|x} \stackrel{\rm JM}{=} \sum_\lambda P(a|x,\lambda)G_\lambda\quad\forall\,a,x.
\end{align}
Namely, a jointly-measurable measurement assemblage can be reproduced (or, say, simulated) by a single measurement (see also Fig.~\ref{Fig:incomp}).
A measurement assemblage is called {\em incompatible} if it is not jointly-measurable.
A profound feature of quantum theory is that there exist incompatible measurements---for instance, ${\bf E}^{\rm Pauli}$ is incompatible.

\subsection{Incompatibility preservability of quantum channels}
To quantify how noisy dynamics affect quantum systems' incompatibility signatures, we still need to know how to model dynamics.
The {\em deterministic} dynamics of quantum systems are described by {\em channels}, which are {\em completely-positive trace-preserving linear maps}~\cite{QIC-book}.
Formally, a channel $\mathcal{N}$ acting on a given system with an input state $\rho$ describes the evolution $\rho~\mapsto~\mathcal{N}(\rho)$.
The {\em stochastic} dynamics of quantum systems are described by {\em filters}, which are {\em completely-positive trace-non-increasing linear maps}.
For an input state $\rho$, a filter $\mathcal{F}$ describes the process $\rho\mapsto\rho/{\rm tr}[\mathcal{F}(\rho)]$ with the success probability ${\rm tr}[\mathcal{F}(\rho)]$.
This allows us to consider stochastic processes (e.g., post-selection) in quantum-information tasks.

\begin{figure}
\begin{center}
\scalebox{0.8}{\includegraphics{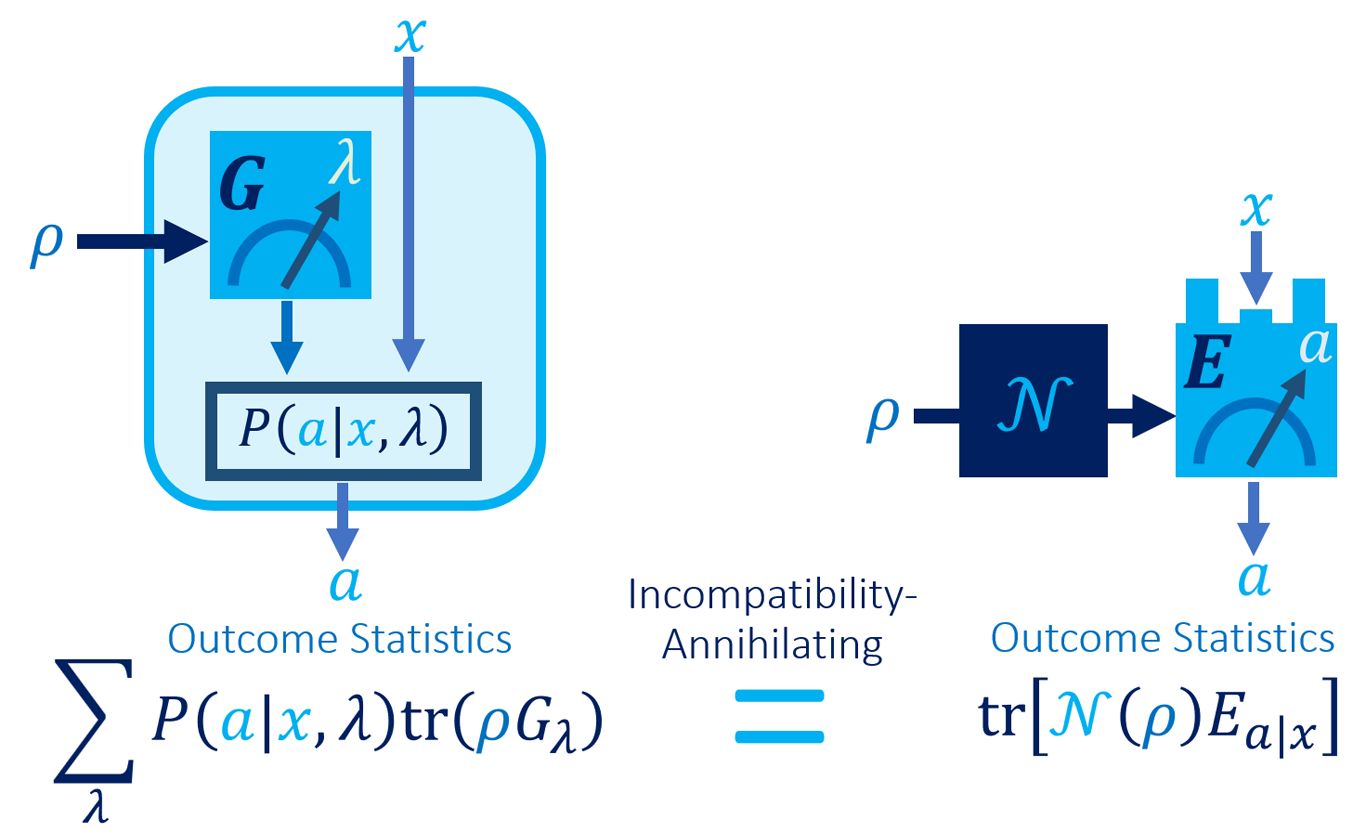}}
\caption{{\bf Incompatibility-annihilating channels.}
A channel $\mathcal{N}$ is incompatibility-annihilating if, after applying it, post-measurement statistics can always be reproduced by jointly-measurable measurements---incompatibility signature is gone.}
\label{Fig:IA} 
\end{center}
\end{figure}

After knowing how to model dynamics, we now formally define those dynamics that {\em cannot} preserve incompatibility for {\em any} input state, \CYnew{as follows:
\begin{definition}
{\em 
A channel $\mathcal{N}$ is said to be {\em incompatibility-annihilating} if, for every measurement assemblage \mbox{${\bf E}\coloneqq\{E_{a|x}\}_{a,x}$,} there exists a jointly-measurable measurement assemblage ${\bf M}^{\rm JM}\coloneqq\{M_{a|x}^{\rm JM}\}_{a,x}$ such that
\begin{align}\label{Eq:def}
{\rm tr}\left[\mathcal{N}(\rho)E_{a|x}\right] = 
{\rm tr}\left(\rho M_{a|x}^{\rm JM}\right)\quad\forall\,\rho,a,x.
\end{align}
}
\end{definition}}
Namely, after applying $\mathcal{N}$ on the system, {\em all possible} measurement statistics can be reproduced by jointly-measurable measurement assemblages (see also Fig.~\ref{Fig:IA})---one can {\em no longer} detect incompatibility.
Crucially, being incompatibility-annihilating is a property of channels {\em independent} of the states they act on; that is, it is a genuinely dynamical property. 
Collectively, we use ${\bf IA}$ to denote the set of all incompatibility-annihilating channels~\footnote{We define ${\bf IA}$ with a {\em fixed} system dimension.}.
Then, \CYnew{we further make the following definition:
\begin{definition}
{\em
A channel $\mathcal{N}$ is said to have {\em incompatibility preservability} if $\mathcal{N}\notin{\bf IA}$.
}
\end{definition}
}
Notably, this is the {\em first} resource preservability theory for preserving resources in the {\em Heisenberg picture}, as all existing ones~\cite{SaxenaPRR2020,Hsieh2020PRR,ChenPRA2021,Hsieh2020Quantum,Stratton2024PRL,Hsieh2021PRXQ} focus on preserving state resources (i.e., in the Schr\"odinger picture).

Importantly, to justify the name ``preservability'', it is crucial to note that $\mathcal{N}$ cannot {\em generate} incompatibility.
This is because Eq.~\eqref{Eq:def} is always true if ${\bf E}$ is already jointly-measurable.
This can be seen by the fact that $\{\mathcal{N}^\dagger(E_{a|x})\}_{a,x}$ is a jointly-measurable measurement assemblage if ${\bf E}$ is jointly-measurable~\cite{Otfried2023RMP,Buscemi2020PRL}.
Hence, when we observe incompatibility after the channel $\mathcal{N}$, it can only be {\em preserved} by $\mathcal{N}$, never generated.
This shows the vital difference between the current framework and resource preservability theories for state resources~\cite{SaxenaPRR2020,Hsieh2020PRR,ChenPRA2021,Hsieh2020Quantum,Stratton2024PRL,Hsieh2021PRXQ}---if we observe state resources from a channel's output, it could be generated (rather than preserved), e.g.,  by a state-preparation channel.

It is worth mentioning that an alternative way to study how a channel removes the incompatibility signature is to use the so-called {\em incompatibility-breaking} channels~\cite{Heinosaari2015,Kiukas2017PRA,Ku2022PRXQ,Ku2023PRA}.
They are channels mapping every measurement assemblage to jointly-measurable ones in the Heisenberg picture.
A natural question is: {\em What is the relation between incompatibility-annihilating channels and incompatibility-\CY{breaking} channels?}
If \mbox{$\mathcal{N}\in{\bf IA}$}, then
\mbox{$
{\rm tr}\left(\rho\left[\mathcal{N}^\dagger(E_{a|x})-M_{a|x}^{\rm JM}\right]\right)=0
$
$\forall\,a,x,\rho$,} meaning that $\mathcal{N}^\dagger(E_{a|x})=M_{a|x}^{\rm JM}$ $\forall\,a,x$, which is exactly the definition of incompatibility-breaking channels; see, e.g., Eq.~(13) in Ref.~\cite{Ku2022PRXQ}.
We thus obtain
\begin{fact}\label{lemma}
$\mathcal{N}\in{\bf IA}$ if and only if  $\{\mathcal{N}^\dagger(E_{a|x})\}_{a,x}$ is a jointly-measurable measurement assemblage for every ${\bf E}$.
Hence, incompatibility-annihilating channels and incompatibility-breaking channels are equivalent.
\end{fact}
While we start with a resource-preservability theory, the underlying channel resource equals an existing one relevant in semi-device-independent tasks~\cite{Ku2022PRXQ}. Hence, our results, once obtained, can provide immediate applications to these operational tasks (we will come back to this later).
Finally, Fact~\ref{lemma} implies ${\bf IA}$ is convex and compact (Lemma~\ref{lemma:convex and compact} in Appendix A).

\section{Formulating the dynamical resource theory of incompatibility preservability}

\begin{figure}
\begin{center}
\scalebox{0.8}{\includegraphics{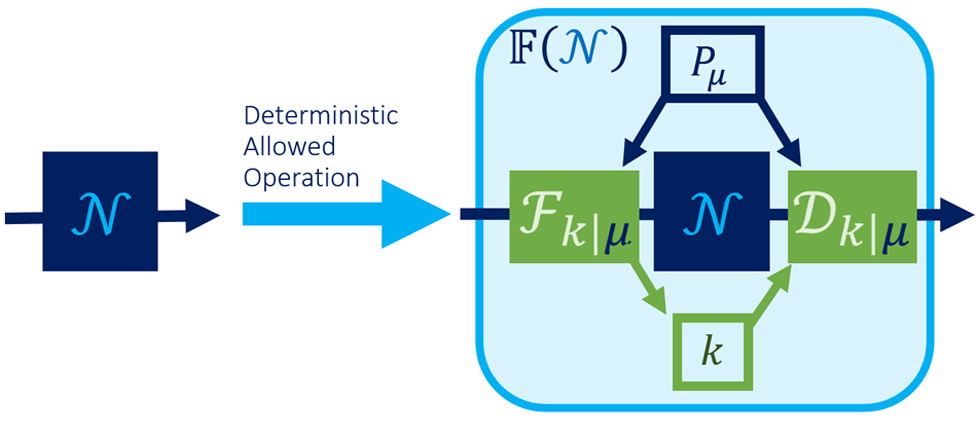}}
\caption{{\bf Schematic illustration of allowed operations.}
For a given channel $\mathcal{N}$, an allowed operation $\mathbb{F}$ maps it to another channel $\mathbb{F}(\mathcal{N})$ (the light blue box) as follows:
First, with probability $P_\mu$, one use the $\mu$-th instrument $\{\mathcal{F}_{k|\mu}\}_k$ and the $\mu$-th set of post-processing channels $\{\mathcal{D}_{k|\mu}\}_k$.
Before applying $\mathcal{N}$, one first applies the instrument, obtaining classical output $k$ and the quantum output $\mathcal{F}_{k|\mu}(\rho)/{\rm tr}[\mathcal{F}_{k|\mu}(\rho)]$ with probability ${\rm tr}[\mathcal{F}_{k|\mu}(\rho)]$.
Then, one applies the given channel $\mathcal{N}$ and the $k$-th channel $\mathcal{D}_{k|\mu}$.
For this $\mu$, on average, we obtain the output state \mbox{$\sum_k{\rm tr}[\mathcal{F}_{k|\mu}(\rho)](\mathcal{D}_{k|\mu}\circ\mathcal{N})(\mathcal{F}_{k|\mu}(\rho)/{\rm tr}[\mathcal{F}_{k|\mu}(\rho)])$}.
Averaging over all possible $\mu$ gives the form of allowed operations [Eq.~\eqref{Eq:DAO}].}
\label{Fig:DAO} 
\end{center}
\end{figure}

\CYnew{\subsection{Allowed operations of incompatibility preservability}}
To start with, a {\em dynamical resource theory} (or {\em channel resource theory}; see, e.g., Refs.~\cite{Navascues2015PRL,RossetPRX2018,Theurer2019PRL,Liu2019,Liu2020PRR,Takagi2020PRL,Hsieh2020Quantum,Hsieh2021PRXQ,Regula2021PRL,Regula2021NC,Hsieh-thesis,Stratton2024PRL}) is defined as a pair of sets. The first set contains channels without the given resource---called {\em free channels}. The second set contains ways in which one can manipulate a channel---called {\em allowed operations}~\footnote{Also known as {\em free operations} in the literature.}.
In the above language, it is clear that incompatibility preservability's free channels are incompatibility-annihilating channels, i.e., the set ${\bf IA}$.
It remains to specify the set of allowed operations to complete the formulation of the resource theory.
To manipulate a channel $\mathcal{N}$, we allow pre-and post-processing quantum operations, and we further allow coordinating the pre-and post-processing parts.
Formally, it is described by some pre-processing filters ($\mathcal{F}_k$ for $k=1,...,|{\bf k}|$) and post-processing channels based on the filtering outcome ($\mathcal{D}_k$ for $k=1,...,|{\bf k}|$), resulting in $\mathcal{N}\mapsto\sum_{k=1}^{|{\bf k}|}\mathcal{D}_k\circ\mathcal{N}\circ\mathcal{F}_k$.
Crucially, as the whole process is deterministic, $\sum_{k=1}^{|{\bf k}|}\mathcal{D}_k\circ\mathcal{N}\circ\mathcal{F}_k$ should be a channel.
Hence, we demand that $\sum_{k=1}^{|{\bf k}|}\mathcal{F}_k$ is a channel, i.e., the set $\{\mathcal{F}_k\}_{k=1}^{|{\bf k}|}$ is an {\em instrument}~\cite{Davies1970CMP}---for an input state $\rho$, it describes the process that, with probability ${\rm tr}[\mathcal{F}_k(\rho)]$, one obtains the classical output index $k$ and quantum output state $\mathcal{F}_k(\rho)/{\rm tr}[\mathcal{F}_k(\rho)]$.
Namely, an instrument can describe a measurement process with both quantum and classical outcomes.
Finally, we also allow using classical randomness to mix different transformations defined above.
Overall, we have a mapping $\mathbb{F}$ as (\CYnew{see Fig.~\ref{Fig:DAO}} for a schematic illustration):
\begin{align}\label{Eq:DAO}
\mathcal{N}\mapsto\mathbb{F}(\mathcal{N})\coloneqq\sum_\mu p_\mu\sum_{k=1}^{|{\bf k}|}\mathcal{D}_{k|\mu}\circ\mathcal{N}\circ\mathcal{F}_{k|\mu},
\end{align}
where $\{p_\mu\}_\mu$ is a probability distribution, $\mathcal{D}_{k|\mu}$'s are channels, and $\mathcal{F}_{k|\mu}$'s are filters such that $\sum_{k=1}^{|{\bf k}|}\mathcal{F}_{k|\mu}$ is a channel $\forall\,\mu$.
Note that $k$'s range can be the same (i.e., $|{\bf k}|$) for different $\mu$'s by putting zero filters when defining $\mathcal{F}_{k|\mu}$'s.
Mappings of this kind are {\em incompatibility preservability's deterministic allowed operations}. Collectively, ${\bf AO}$ denotes the set of all mappings defined via Eq.~\eqref{Eq:DAO}.

Our justification of choosing ${\bf AO}$ is that, in addition to being physically implementable in labs, they cannot generate incompatibility preservability from an incompatibility-annihilating channel, sometimes called the {\em golden rule} of resource theories~\cite{ChitambarRMP2019} (see Appendix A for proof):
\begin{theorem}\label{Result1}
$\mathbb{F}(\mathcal{N})\in{\bf IA}$ for every $\mathcal{N}\in{\bf IA}$ and $\mathbb{F}\in{\bf AO}$.
\end{theorem}
Hence, if a system undergoes an incompatibility-annihilating channel $\mathcal{N}$, then no operation $\mathbb{F}$ from ${\bf AO}$ can prevent the system's incompatibility signature from being destroyed.
In particular, applying filters $\mathcal{F}_{k|\mu}$'s {\em before} the incompatibility-annihilating channel $\mathcal{N}$ {\em cannot} help, as long as the whole operation $\mathbb{F}$ is deterministic.

Interestingly, ${\bf AO}$ coincides with the allowed operations of quantum memory~\cite{RossetPRX2018}, whose free channels are the so-called entanglement-breaking channels~\cite{Horodecki2003RevMP}.
Hence, although entanglement-breaking and incompatibility-annihilating channels are essentially different dynamical properties (since being entanglement-breaking {\em does not equal} being incompatibility-breaking~\cite{Ku2022PRXQ}), their resource theories can have the same allowed operations---both quantum memory and incompatibility preservability cannot be generated by the same class of manipulations, i.e., ${\bf AO}$~\footnote{${\bf AO}$ does not equal the largest sets of allowed operations for quantum memory and incompatibility preservability. The relation between these two sets is still unknown.}.

\subsection{Quantifying incompatibility preservability via operational tasks}
\begin{figure}
\begin{center}
\scalebox{0.8}{\includegraphics{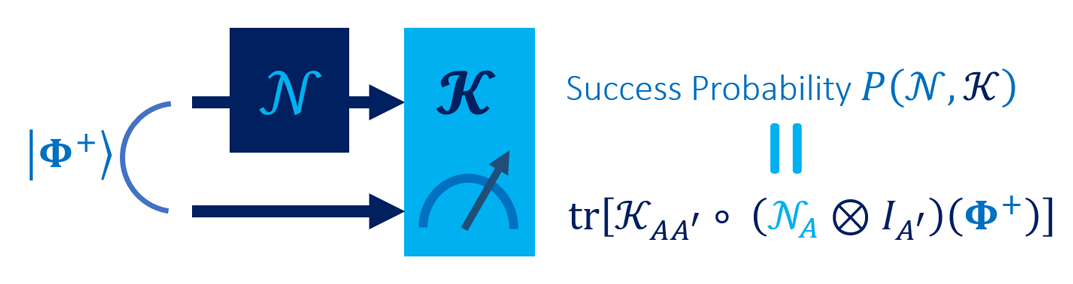}}
\caption{{\bf Entanglement-assisted filter games.}
\CY{Player locally applies} a channel $\mathcal{N}$ to
the maximally entangled state $\Phi^+\coloneqq\proj{\Phi^+}_{AA'}$, resulting in the score $P(\mathcal{N},\mathcal{K})$, which is \CY{$\mathcal{K}$'s} success probability.}
\label{Fig:game} 
\end{center}
\end{figure}

Now, we quantify incompatibility preservability via its dynamical resource theory.
To this end, we consider the {\em incompatibility preservability robustness} $\mathbfcal{R}$ defined as
\begin{align}\label{Eq:robustness}
\mathbfcal{R}(\mathcal{N})\coloneqq\min\left\{t\ge0\,\middle|\,\frac{\mathcal{N}+t\mathcal{W}}{1+t}\in{\bf IA},\;\text{$\mathcal{W}$: channel}\right\}.
\end{align}
Now, as proved in Appendix B, we show that:
\begin{theorem}\label{Result:monotone}
$\mathbfcal{R}(\mathcal{N})\ge0$ for every channel $\mathcal{N}$, and the equality holds if and only if $\mathcal{N}\in{\bf IA}$.
Moreover, 
\begin{align}
\text{$\mathbfcal{R}[\mathbb{F}(\mathcal{N})]\le\mathbfcal{R}(\mathcal{N})$\quad$\forall$ channel $\mathcal{N}$ and $\mathbb{F}\in{\bf AO}$.}
\end{align} 
\end{theorem}
Thus, $\mathbfcal{R}$ is monotonic under allowed operations, and one can view it as a valid {\em quantifier} of incompatibility preservability.
In fact, $\mathbfcal{R}$ has a clear operational characterisation, which can help us quantify incompatibility preservability via an operational task termed {\em entanglement-assisted filter game}.
The game starts with a bipartite system $AA'$ with equal local dimension $d<\infty$. 
A bipartite maximally entangled state is prepared as the input: 
\CYnew{
\begin{align}
\ket{\Phi^+}_{AA'} \coloneqq \sum_{n=0}^{d-1}\ket{nn}_{AA'}/\sqrt{d}.
\end{align}}
\CYnew{Also,} since now, whenever needed, subscripts denote the systems the operators act on.
To define the game, a referee will fix \CY{a filter $\mathcal{K}$ in $AA'$.}
In the game, the ``score'' is \CY{$\mathcal{K}$'s} success probability. For a better score, the player can choose a channel $\mathcal{N}$ (i.e., a ``strategy'') acting on $A$ before the filter. 
The player's score then reads (see also Fig.~\ref{Fig:game}):
\begin{align}\label{Eq:score}
P(\mathcal{N},\mathcal{K})\coloneqq{\rm tr}\left[\CY{\mathcal{K}_{AA'}\circ}(\mathcal{N}_A\otimes\mathcal{I}_{A'})(\proj{\Phi^+}_{AA'})\right].
\end{align}
\CY{When $\mathcal{K}$ is in the \CYnew{following specific} form 
\CYnew{
\begin{align}
\mathcal{K}(\cdot) = \sqrt{K}(\cdot)\sqrt{K}
\end{align}}
for some $0\le K\le\id$, we particularly call it an {\em ${\rm F_1}$} filter}~\footnote{We choose this name since {\em local} filters with a single Kraus operator are called {\em ${\rm LF_1}$ filters} in the literature~\cite{Ku2022NC,ku2023,Hsieh2023}.} (see, e.g., Refs.~\cite{Ku2022NC,ku2023,Hsieh2023,Hsieh2024,Tabia2024}, for its applications).
It turns out that this task provides a clear operational interpretation of $\mathbfcal{R}$.
In Appendix C, we prove that
\begin{theorem}\label{Result:operational meaning}
For every channel $\mathcal{N}$, we have
\begin{align}\label{Eq:Result:operational meaning01}
1+\mathbfcal{R}(\mathcal{N}) = \max_{\mathcal{K}:{\rm F_1}}\frac{P(\mathcal{N},\mathcal{K})}{\displaystyle\max_{\mathcal{L}\in{\bf IA}}P(\mathcal{L},\mathcal{K})}\CY{= \max_{\mathcal{K}}\frac{P(\mathcal{N},\mathcal{K})}{\displaystyle\max_{\mathcal{L}\in{\bf IA}}P(\mathcal{L},\mathcal{K})}},
\end{align}
where ``$\max_{\mathcal{K}:{\rm F_1}}$'' maximises over all ${\rm F_1}$ filters, \CY{and ``$\max_{\mathcal{K}}$'' maximises over all filters}.
\end{theorem}
\CY{Hence,} the information-theoretical quantity $\mathbfcal{R}$ \CY{has} a clear physical meaning in an operational task---$\mathbfcal{R}$ is the \CY{optimal} advantage in the filter game, and incompatibility preservability can provide higher filter success probabilities.
\CY{Theorem~\ref{Result:operational meaning} also uncovers a no-go result---general filters cannot outperform $F_1$ filters in maximising the ratio given in Eq.~\eqref{Eq:Result:operational meaning01}.
Hence, to practically evaluate $\mathbfcal{R}$, it suffices to implement ${\rm F_1}$ filters, which are experimentally friendly compared with general filters.}

\subsection{Certifying incompatibility preservability by channel's nonlocal resources}
\CY{A physical implication of Theorem~\ref{Result:operational meaning} is that channels' ability to keep nonlocal resources can certify incompatibility preservability.
To see this, consider the ${\rm F_1}$ filter \mbox{$\mathcal{K}(\cdot) = \sqrt{K}(\cdot)\sqrt{K}$} with} $K=\proj{\Phi^+}$.
Using \CY{the} steering bound induced by mutually unbiased basis~\cite{Hsieh2016PRA}, for $d$ as a power of prime number, gives (see Appendix C):
\begin{align}\label{Eq:MUB R}
1+\mathbfcal{R}(\mathcal{N})\ge\mathbfcal{F}_+(\mathcal{N}) d\sqrt{d+1}/(d-1+\sqrt{d+1}),
\end{align}
where 
\begin{align}
\mathbfcal{F}_+(\mathcal{N})\coloneqq\bra{\Phi^+}(\mathcal{N}_A\otimes\mathcal{I}_{A'})(\proj{\Phi^+}_{AA'})\ket{\Phi^+}
\end{align}
is \CY{$\mathcal{N}$'s} {\em singlet fraction}~\cite{Buscemi2010ITIT,Horodecki1999PRA}. 
\CY{A bipartite state $\rho_{AA'}$'s singlet fraction, $\bra{\Phi_+}\rho_{AA'}\ket{\Phi_+}$, is well-known for measuring its nonlocal resources, e.g., entanglement~\cite{Horodecki2009RMP,Hsieh2020PRR}, steering~\cite{UolaRMP2020,RennerPRL2024,ZhangPRL2024,Hsieh2016PRA}, nonlocality~\cite{Brunner2014RMP,Palazuelos2012PRL}, usefulness of teleportation~\cite{Horodecki1999PRA,Li2021PRR,CavalcantiPRA2013}. 
Thus, for a channel acting on $A$, $\mathbfcal{F}_+$ measures its ability to keep nonlocal resources shared between $A$ and an external system $A'$}.
\CY{Equation~\eqref{Eq:MUB R} implies} 
\CYnew{
\begin{corollary}
$\mathcal{N}$ has incompatibility preservability if
\begin{align}
\mathbfcal{F}_+(\mathcal{N})>(d-1+\sqrt{d+1})/d\sqrt{d+1}.
\end{align}
\end{corollary}}
\CY{Hence, a strong enough ability to keep nonlocal resources (in $AA'$) {\em implies} the ability to preserve incompatibility (in $A$).
Notably, in the large system limit}
$d\gg1$, the bound reduces to $\mathbfcal{F}_+(\mathcal{N})\gtrsim1/\sqrt{d}$.
\CY{Also,} when $d=2$, \CY{the} exact steering bound~\cite{RennerPRL2024,ZhangPRL2024} gives \CY{(see Appendix C)}
\begin{align}\label{Eq:qubit R}
1+\mathbfcal{R}(\mathcal{N})\ge\mathbfcal{F}_+(\mathcal{N})8/5.
\end{align}
Namely, \CYnew{we have
\begin{corollary}
A qubit channel $\mathcal{N}$ has incompatibility preservability if
\begin{align}
\mathbfcal{F}_+(\mathcal{N})>5/8.
\end{align}
\end{corollary}}
Hence, \CY{interestingly, when channels have strong enough abilities to keep nonlocal resources (measured by $\mathbfcal{F}_+$), they must also have incompatibility preservability.}

\subsection{Complete characterisation of incompatibility preservability conversion}
To complete the resource theory, we now fully characterise resource conversion via the entanglement-assisted filter games.
In the game, suppose the player already decides to use the channel $\mathcal{N}$. 
Then, by using ${\bf AO}$ to further manipulate $\mathcal{N}$, the player can enhance the score Eq.~\eqref{Eq:score} to
\begin{align}
P_{\rm max}(\mathcal{N},\mathcal{K})\coloneqq\max_{\mathbb{F}\in{\bf AO}}P[\mathbb{F}(\mathcal{N}),\mathcal{K}].
\end{align}
Similar to the case of quantification, $P_{\rm max}$ can fully characterise the resource conversion (see Appendix D for proof):
\begin{theorem}\label{Result:conversion}
Let $\mathcal{N},\mathcal{M}$ be two channels.
The following three statements are equivalent:
\begin{enumerate}
\item $\mathbb{F}(\mathcal{N}) = \mathcal{M}$ for some $\mathbb{F}\in{\bf AO}$.
\item $P_{\rm max}(\mathcal{N},\mathcal{K})\ge P_{\rm max}(\mathcal{M},\mathcal{K})$ for every ${\rm F_1}$ filter $\mathcal{K}$.
\item \CY{$P_{\rm max}(\mathcal{N},\mathcal{K})\ge P_{\rm max}(\mathcal{M},\mathcal{K})$ for every filter $\mathcal{K}$.}
\end{enumerate}
\end{theorem}
Hence, if $\mathcal{N}$ admits a score $P_{\rm max}$ no less than $\mathcal{M}$'s in {\em every} game $\mathcal{K}$, it is {\em equivalent to} being able to convert $\mathcal{N}$ to $\mathcal{M}$ via some allowed operations.
Importantly, the above result can be generalised to {\em arbitrary channel resource theories} as long as the set of deterministic allowed operations is convex, compact, containing the identity map, and is closed under function composition
(see Theorem~\ref{Result: general DRT conversion} in Appendix D).
Hence, as a bonus, the entanglement-assisted filter games can completely characterise {\em general} channel resources.
\CY{Finally, Theorem~\ref{Result:conversion} suggests that ${\rm F_1}$ filters are {\em complete} for characterising resource conversion.
This is of particular practical value, as ${\rm F_1}$ filters are much more experimentally feasible compared with general ones.}

\subsection{Applications to semi-device-independent resources}
Interestingly, as an application, our framework also {\em completes} the resource theories for a recently-proposed semi-device-independent channel resource, termed {\em non-steering-breaking} channels~\cite{Ku2022PRXQ}.
They are local channels preserving quantum steering of globally distributed maximally entangled states (Theorem 1 in Ref.~\cite{Ku2022PRXQ})---they maintain global quantum signatures semi-device-independently.
Since incompatibility-breaking channels are also steering-breaking (Theorem 2 in Ref.~\cite{Ku2022PRXQ}), Fact~\ref{lemma} implies that a channel has incompatibility preservability if and only if it is non-steering-breaking.
Consequently, incompatibility preservability's resource theory can also be the resource theory for non-steering-breaking channels, which lacks a complete resource theory until now.
Due to this observation, the robustness $\mathbfcal{R}$ equals the non-steering breaking robustness and the non-incompatibility-breaking robustness (Eqs.~(14) and~(15) in Ref.~\cite{Ku2022PRXQ}).
Our results thus show that these two existing robustness measures are monotonic under ${\bf AO}$ (Theorem~\ref{Result:monotone}) and have a clear operational meaning (Theorem~\ref{Result:operational meaning}).
Finally, Theorem~\ref{Result:conversion} also gives a necessary and sufficient characterisation of converting non-steering-breaking and non-incompatibility-breaking channels.
Hence, our results have immediate applications to semi-device-independent scenarios via its equivalence with non-steering-breaking channels~\cite{Ku2022PRXQ}.

\subsection{Illustrative examples}
Finally, as an example, consider the {\em depolarising channel} $\Lambda_p(\cdot)\coloneqq p\mathcal{I}(\cdot) + (1-p){\rm tr}(\cdot)\id/d$ in a $d$-dimensional system, where $\mathcal{I}$ is the identity channel.
Note that $\Lambda_p\in{\bf IA}$ if and only if it is incompatibility-breaking, which is true if and only if it is steering-breaking~\cite{Ku2022PRXQ}.
Utilising the steering bound~\cite{Wiseman2007PRL} (see also Eq.~(55) in Ref.~\cite{UolaRMP2020}), we obtain $\Lambda_p\in{\bf IA}$ if \mbox{$p\le(H_d-1)/(d-1)$}, where $H_d = 1+1/2+...+1/d$ (the ``only if'' holds for $d=2$~\cite{RennerPRL2024,ZhangPRL2024}).
Hence, if we depolarise the system via ${\rm tr}(\cdot)\id/d$ with probability $1-p$, we can remove incompatibility signature when $p$ is small enough.
More precisely, \CYnew{we have the following result:}
\CYnew{\begin{proposition}
for every channel $\mathcal{N}$ and measurement assemblage ${\bf E}$, there is a jointly-measurable ${\bf M}^{\rm JM}$ such that 
\begin{align}
{\rm tr}(E_{a|x}\mathcal{N}[\Lambda_p(\rho)]) = {\rm tr}(M_{a|x}^{\rm JM}\rho)\quad
\forall\,a,x,\rho,
\end{align}
if \mbox{$p\le(H_d-1)/(d-1)$}.
\end{proposition}}
Thus, if a system goes through $\Lambda_p$ with a small $p$, measurement statistics can {\em always} be reproduced by jointly-measurable measurement assemblages.
Finally, when $d=2$, we can use Eq.~\eqref{Eq:qubit R}, the fact that \mbox{$\mathbfcal{F}_+(\Lambda_p)=(3p+1)/4$,} and the exact steering bound~\cite{RennerPRL2024,ZhangPRL2024} to obtain
\begin{align}
\max\left\{0;(2p-1)\right\}\times3/5\le\mathbfcal{R}(\Lambda_p)\le\max\{0;2p-1\}.
\end{align}
This provides an explicit formula for bounding $\mathbfcal{R}(\Lambda_p)$.
We leave further explorations of our framework to future projects.

\CYnew{\section{Conclusion}
We introduce the resource theory of incompatibility preservability to quantify and characterise dynamics' ability to preserve incompatibility signature. Our framework is the first resource preservability theory for quantum resources in the Heisenberg picture, providing a platform for future investigations on preserving resources of measurements.
}

Several questions remain open.
First, further exploring our framework's applications to semi-device-independent scenarios is of great value.
Second, it is interesting to study whether exclusion-type operational tasks~\cite{Ducuara2020PRL,Ducuara2022PRXQ,Ducuara2023PRL,stratton2024,Hsieh-IP} can quantify/characterise incompatibility preservability.
Third, it is still unknown whether our framework can be naturally extended to understanding the preservation of incompatibility of instruments~\cite{Mitra2022PRA,Mitra2023PRA,Ji2024PRXQ,Buscemi2023Quantum,Leppajarvi2024incompatibilityof,Hsieh2024,Ji2024PRXQ}, channels~\cite{Haapasalo2021,Hsieh2022PRR}, and marginal problems~\cite{Hsieh2024Quantum,Tabia2022npjQI}.
Finally, it will be valuable to uncover any thermodynamic applications of incompatibility preservability via the recently found connections between incompatibility, steering, and thermodynamics~\cite{Hsieh2024,Hsieh2024Quantum,JiPRL2022,BeyerPRL2019,ChanPRA2022}.

\subsection*{Acknowledgements}
The authors acknowledge fruitful discussions with Paul Skrzypczyk.
\CY{We also thank the anonymous referee for pointing out that Theorems~\ref{Result:operational meaning},~\ref{Result:conversion}, and~\ref{Result: general DRT conversion} actually hold for general filters, and consequently imply the no-go results presented in this work.
C.-Y.~H. acknowledges support from the Royal Society through Enhanced Research Expenses (on grant NFQI) and the Leverhulme Trust
Early Career Fellowship (on grant “Quantum complementarity: a novel resource for quantum science and technologies”
with grant number ECF-2024-310).}
H.-Y.~K. is supported by the
Ministry of Science and Technology, Taiwan, (with grant number MOST 112-2112-M-003-020-MY3), and Higher Education Sprout Project of National Taiwan Normal University (NTNU).
B.~S.~acknowledges support from UK EPSRC (EP/SO23607/1).

\appendix

\section{Proof of Theorem~\ref{Result1}}
First, we establish the following lemma.
Let ${\bf JM}$ be the set of all jointly-measurable measurement assemblages.
\begin{lemma}\label{lemma:convex and compact}
${\bf IA}$ is convex and compact in the diamond norm.
\end{lemma}
{\em Proof.} Since ${\bf JM}$ is convex, ${\bf IA}$'s convexity follows.
It remains to prove ${\bf IA}$'s compactness in the metric topology induced by the {\em diamond norm}~~\cite{Aharonov1998,watrous_2018}, $\norm{\cdot}_\diamond$, which is defined for two channels $\mathcal{N},\mathcal{M}$ acting on a system $S$ as 
\begin{align}
\CYnew{\norm{\mathcal{N} - \mathcal{M}}_\diamond\coloneqq\sup_{\rho_{SA}}\norm{\left[(\mathcal{N} - \mathcal{M})_S\otimes\mathcal{I}_A\right](\rho_{SA})}_1.}
\end{align}
It maximises over every possible bipartite state $\rho_{SA}$ and (finite-dimensional) auxiliary system $A$. 
Since ${\bf IA}$ is a strict subset of the set of all channels, it is bounded. Hence, it suffices to show that ${\bf IA}$ is a closed set to conclude its compactness.
To this end, we will show that ${\bf IA}$ contains its closure. 
Suppose $\mathcal{L}$ is in ${\bf IA}$'s closure.
Then there exists a sequence $\{\mathcal{N}_k\}_{k=1}^\infty\subset{\bf IA}$ such that $\lim_{k\to\infty}\norm{\mathcal{N}_k - \mathcal{L}}_\diamond=0$.
Since $\mathcal{N}_k\in{\bf IA}$ $\forall\,k$, Fact~\ref{lemma} implies that, for any $\{E_{a|x}\}_{a,x}$ and $k$, the set $\{\mathcal{N}_k^\dagger(E_{a|x})\}_{a,x}\in{\bf JM}$.
Now, $\lim_{k\to\infty}\norm{\mathcal{N}_k - \mathcal{L}}_\diamond=0$ implies~\footnote{\CY{Since we only consider finite-dimensional systems, both sup norm ``$\norm{\cdot}_\infty$'' and trace norm ``$\norm{\cdot}_1$'' induce the same topology and are interchangeable. We thus use trace norm here due to its connection to the diamond norm, while we note that the sup norm is a more natural measure in the Heisenberg picture.}} 
\begin{align}
\CYnew{\lim_{k\to\infty}\norm{\mathcal{N}_k^\dagger(E_{a|x}) - \mathcal{L}^\dagger(E_{a|x})}_1=0
\quad
\forall\,a,x.}
\end{align}
Hence, $\{\mathcal{L}^\dagger(E_{a|x})\}_{a,x}$ is in ${\bf JM}$'s closure.
Since ${\bf JM}$ is compact, it contains its closure, meaning that 
\begin{align}
\CYnew{
\{\mathcal{L}^\dagger(E_{a|x})\}_{a,x}\in{\bf JM}.
}
\end{align}
Since this argument works for {\em any} given measurement assemblage $\{E_{a|x}\}_{a,x}$, Fact~\ref{lemma} implies that $\mathcal{L}\in{\bf IA}$, i.e., ${\bf IA}$ contains its closure and is thus a closed set.
\hfill $\square$

Now, we are in the position to prove Theorem~\ref{Result1}:

{\em Proof of Theorem~\ref{Result1}.}
Consider $\mathcal{N}\in{\bf IA}$ and $\mathbb{F}\in{\bf AO}$.
Write $\mathbb{F}(\mathcal{N}) = \sum_{\mu,k} p_\mu\mathcal{D}_{k|\mu}\circ\mathcal{N}\circ\mathcal{F}_{k|\mu}$.
Let $\{E_{a|x}\}_{a,x}$ be a given measurement assemblage.
Then, for every state $\rho$, 
\begin{align}
{\rm tr}\left[\mathbb{F}(\mathcal{N})(\rho)E_{a|x}\right]=\sum_{\mu,k}p_\mu{\rm tr}\left[\mathcal{N}\left(\mathcal{F}_{k|\mu}(\rho)\right)\mathcal{D}_{k|\mu}^\dagger(E_{a|x})\right].
\end{align}
For every $\mu,k$, $\{\mathcal{D}_{k|\mu}^\dagger(E_{a|x})\}_{a,x}$ is a valid measurement assemblage. Since $\mathcal{N}\in{\bf IA}$, Eq.~\eqref{Eq:def} implies that there is a measurement assemblage $\{M_{a|x}^{(k|\mu)}\}_{a,x}\in{\bf JM}$ such that [``$(\cdot)$'' means that it is for every possible input state]
\begin{align}
{\rm tr}\left[\mathcal{N}\left(\cdot\right)\mathcal{D}_{k|\mu}^\dagger(E_{a|x})\right] = {\rm tr}\left[(\cdot)M_{a|x}^{(k|\mu)}\right]\quad\forall\,a,x.
\end{align}
Hence, for every $k,\mu$, we substitute the state $\mathcal{F}_{k|\mu}(\rho)/{\rm tr}[\mathcal{F}_{k|\mu}(\rho)]$ into ``$(\cdot)$'' and obtain (without loss of generality, we assume ${\rm tr}[\mathcal{F}_{k|\mu}(\rho)]>0$ here)
\begin{align}
{\rm tr}\left[\mathcal{N}\left(\mathcal{F}_{k|\mu}(\rho)\right)\mathcal{D}_{k|\mu}^\dagger(E_{a|x})\right] = {\rm tr}\left[\rho\mathcal{F}_{k|\mu}^\dagger\left(M_{a|x}^{(k|\mu)}\right)\right]\quad\forall\,a,x.
\end{align}
Since the above argument works for every $\rho$, we conclude that
\begin{align}\label{Eq:Computation001}
{\rm tr}\left[\mathbb{F}(\mathcal{N})(\rho)E_{a|x}\right] = {\rm tr}\left[\rho\sum_{\mu,k}p_\mu\mathcal{F}_{k|\mu}^\dagger\left(M_{a|x}^{(k|\mu)}\right)\right]\quad\forall\,a,x,\rho.
\end{align}
Now, define 
\begin{align}
\CYnew{
W_{a|x}\coloneqq\sum_{\mu,k}p_\mu\mathcal{F}_{k|\mu}^\dagger\left(M_{a|x}^{(k|\mu)}\right)
\quad
\forall\,a,x.
}
\end{align}
One can check that $W_{a|x}\ge0$ $\forall\,a,x$ and $\sum_aW_{a|x}=\id$ $\forall\,x$. 
Hence, $\{W_{a|x}\}_{a,x}$ is a valid measurement assemblage.
Now, we will show that it is also jointly-measurable.
To see this, since $\{M_{a|x}^{(k|\mu)}\}_{a,x}\in{\bf JM}$ $\forall\,k,\mu$, we have \mbox{$M_{a|x}^{(k|\mu)} = \sum_iD(a|x,i)G_i^{(k|\mu)}$} for some POVM $\{G_i^{(k|\mu)}\}_i$ and deterministic probability distributions $D(a|x,i)$'s~\cite{Ali2009FP}.
See also Ref.~\cite{Otfried2023RMP}, especially Eqs.~(19-22) and the discussion above them.
This means
\begin{align}
W_{a|x} = \sum_{i}D(a|x,i)\sum_{\mu,k}p_\mu\mathcal{F}_{k|\mu}^\dagger\left(G_i^{(k|\mu)}\right)\quad\forall\,a,x.
\end{align}
Define 
\begin{align}
\CYnew{
\widetilde{G}_i\coloneqq\sum_{\mu,k}p_\mu\mathcal{F}_{k|\mu}^\dagger\left(G_i^{(k|\mu)}\right)
\quad
\forall\,i.
}
\end{align}
They satisfy $\widetilde{G}_i\ge0$ $\forall\,i$ and
\begin{align}
\CYnew{
\sum_i\widetilde{G}_i=\sum_\mu p_\mu\left(\sum_k\mathcal{F}_{k|\mu}^\dagger\right)\left(\id\right)=\id,
}
\end{align}
where we have used the fact that $\sum_k\mathcal{F}_{k|\mu}^\dagger=\left(\sum_k\mathcal{F}_{k|\mu}\right)^\dagger$ is a unital map since $\sum_k\mathcal{F}_{k|\mu}$ is trace-preserving.
Hence, $\{\widetilde{G}_i\}_i$ is a valid POVM.
By definition [Eq.~\eqref{Eq:JM}], $\{W_{a|x}\}_{a,x}\in{\bf JM}$.
Using Eqs.~\eqref{Eq:def} and~\eqref{Eq:Computation001}, we conclude that $\mathbb{F}(\mathcal{N})\in{\bf IA}$.
\hfill $\square$


\section{Proof of Theorem~\ref{Result:monotone}}
Since ${\bf IA}$ is compact (Lemma~\ref{lemma:convex and compact}), $\mathbfcal{R}(\mathcal{N})=0$ if and only if $\mathcal{N}\in{\bf IA}$.
Finally, a direct computation shows that
\begin{align}
\mathbfcal{R}&[\mathbb{F}(\mathcal{N})] = \min\left\{t\ge0\,\middle|\,\frac{\mathbb{F}(\mathcal{N})+t\mathcal{W}}{1+t}\in{\bf IA},\mathcal{W}:\text{channel}\right\}\nonumber\\
&\le\min\left\{t\ge0\,\middle|\,\frac{\mathbb{F}(\mathcal{N})+t\mathbb{F}(\mathcal{W}')}{1+t}\in{\bf IA},\mathcal{W}':\text{channel}\right\}\nonumber\\
&\le\min\left\{t\ge0\,\middle|\,\frac{\mathcal{N}+t\mathcal{W}'}{1+t}\in{\bf IA},\mathcal{W}':\text{channel}\right\}.
\end{align}
The last inequality follows from Theorem~\ref{Result1} and linearity of $\mathbb{F}$; namely, 
\begin{align}
\CYnew{
\frac{\mathbb{F}(\mathcal{N})+t\mathbb{F}(\mathcal{W}')}{1+t}=\mathbb{F}\left(\frac{\mathcal{N}+t\mathcal{W}'}{1+t}\right)\in{\bf IA}
}
\end{align}
if $\CYnew{(\mathcal{N}+t\mathcal{W}')/(1+t)}\in{\bf IA}$.
\hfill$\square$

\section{Proof of Theorem~\ref{Result:operational meaning}}
First, $\mathbfcal{R}$ equals the generalised robustness measure for channels defined in 
Eq.~(54) in Ref.~\cite{Takagi2019PRX} with respect to the set ${\bf IA}$.
Using Theorem 5 and Eq.~(73) in Ref.~\cite{Takagi2019PRX}, there exists a bipartite operator $Y\ge0$ in $AA'$ achieving (let $\Phi^+\coloneqq\proj{\Phi^+}$)
\begin{align}
1+\mathbfcal{R}(\mathcal{N})\le\frac{{\rm tr}\left[Y_{AA'}(\mathcal{N}_A\otimes\mathcal{I}_{A'})(\Phi^+_{AA'})\right]}{\max_{\mathcal{L}\in{\bf IA}}{\rm tr}\left[Y_{AA'}(\mathcal{L}_A\otimes\mathcal{I}_{A'})(\Phi^+_{AA'})\right]}.
\end{align}
Defining the ${\rm F_1}$ filter 
\begin{align}
\CYnew{
\mathcal{Y}(\cdot)\coloneqq \sqrt{Y}(\cdot)\sqrt{Y}/\norm{Y}_\infty
}
\end{align}
and using Eq.~\eqref{Eq:score} give the upper bound
\begin{align}\label{Eq:comp 000001}
1+\mathbfcal{R}(\mathcal{N})&\le\frac{P(\mathcal{N},\mathcal{Y})}{\max_{\mathcal{L}\in{\bf IA}}P(\mathcal{L},\mathcal{Y})}\le\max_{\mathcal{K}:{\rm F_1}}\frac{P(\mathcal{N},\mathcal{K})}{\max_{\mathcal{L}\in{\bf IA}}P(\mathcal{L},\mathcal{K})}\nonumber\\
&\CY{\le\max_{\mathcal{K}}\frac{P(\mathcal{N},\mathcal{K})}{\max_{\mathcal{L}\in{\bf IA}}P(\mathcal{L},\mathcal{K})}.}
\end{align}
\CY{For any filter $\mathcal{K}$, the score Eq.~\eqref{Eq:score} can be re-written as~\footnote{\CY{We thank the anonymous referee for pointing this out.}}
\begin{align}\label{Eq:thx to referee!}
P(\mathcal{N},\mathcal{K})={\rm tr}\left[\Gamma_{AA'}(\mathcal{N}_A\otimes\mathcal{I}_{A'})(\proj{\Phi^+}_{AA'})\Gamma_{AA'}\right],
\end{align}
where \CYnew{the operator}
\begin{align}
\CYnew{
\Gamma_{AA'}\coloneqq\sqrt{\mathcal{K}_{AA'}^\dagger(\id_{AA'})}
}
\end{align}
satisfies the condition \mbox{$\Gamma_{AA'}^2 = \mathcal{K}_{AA'}^\dagger(\id_{AA'})\le\id_{AA'}$,} meaning that $0\le\Gamma_{AA'}\le\id_{AA'}$. Hence, the map 
\begin{align}
\CYnew{(\cdot)\mapsto\Gamma_{AA'}(\cdot)\Gamma_{AA'}}
\end{align}
is an $F_1$ filter, and the last inequality in Eq.~\eqref{Eq:comp 000001} is an equality.
Consequently, it suffices to show the ``$\max_{\mathcal{K}:{\rm F_1}}$'' upper bound in Eq.~\eqref{Eq:comp 000001} can be saturated.
Let \mbox{$t_*=\mathbfcal{R}(\mathcal{N})$}.}
By Eq.~\eqref{Eq:robustness}, there exist $\widetilde{\mathcal{L}}\in{\bf IA}$ and a channel $\widetilde{\mathcal{W}}$ achieving 
\mbox{$
\mathcal{N} + t_*\widetilde{\mathcal{W}} = (1+t_*)\widetilde{\mathcal{L}}.
$}
Hence, for every ${\rm F_1}$ filter \mbox{$\mathcal{K}(\cdot) = \sqrt{K}(\cdot)\sqrt{K}$} \CY{in $AA'$, we have} 
\begin{align}
P(\mathcal{N},\mathcal{K})&\le(1+t_*){\rm tr}[K_{AA'}(\widetilde{\mathcal{L}}_A\otimes\mathcal{I}_{A'})(\Phi^+_{AA'})]\nonumber\\
&\le(1+t_*)\max_{\mathcal{L}\in{\bf IA}}P(\mathcal{L},\mathcal{K}).
\end{align}
Dividing both sides by $\max_{\mathcal{L}\in{\bf IA}}P(\mathcal{L},\mathcal{K})$ and maximising over all ${\rm F_1}$ filters, we obtain
\begin{align}
\max_{\mathcal{K}:{\rm F_1}}\frac{P(\mathcal{N},\mathcal{K})}{\max_{\mathcal{L}\in{\bf IA}}P(\mathcal{L},\mathcal{K})}\le1+\mathbfcal{R}(\mathcal{N}),
\end{align}
which completes the proof.
\hfill$\square$

As an example, by considering \CY{the specific ${\rm F_1}$ filter \mbox{$\mathcal{K}(\cdot) = \sqrt{K}(\cdot)\sqrt{K}$} with} $K = \proj{\Phi^+}$, we obtain
\begin{align}
1+\mathbfcal{R}(\mathcal{N})\ge\frac{\mathbfcal{F}_+(\mathcal{N})}{\max_{\mathcal{L}\in{\bf IA}}\mathbfcal{F}_+(\mathcal{L})}.
\end{align}
Since being incompatibility-annihilating equals being incompatibility-breaking (Fact~\ref{lemma}), which breaks the steerability of $\ket{\Phi^+}$ (Theorems 1 and 2 in Ref.~\cite{Ku2022PRXQ}), we have
\begin{align}
\max_{\mathcal{L}\in{\bf IA}}\mathbfcal{F}_+(\mathcal{L})\le\max_{\eta\in{\bf LHS}}\bra{\Phi^+}\eta_{AA'}\ket{\Phi^+},
\end{align}
where the right-hand side maximises over all bipartite states in $AA'$ admitting local-hidden-state models (see Ref.~\cite{Ku2022PRXQ} for details). 
This provides a useful tool to estimate $\mathbfcal{R}$.
For instance, when $d$ is a power of a prime number, Eq.~(23) in Ref.~\cite{Hsieh2016PRA} implies
\begin{align}
\CYnew{
\max_{\eta\in{\bf LHS}}\bra{\Phi^+}\eta_{AA'}\ket{\Phi^+}\le(d-1+\sqrt{d+1})/d\sqrt{d+1},}
\end{align}
leading to Eq.~\eqref{Eq:MUB R}.
Also, when $d=2$, the exact steering bound~\cite{RennerPRL2024,ZhangPRL2024} implies 
\begin{align}
\CYnew{
\max_{\eta\in{\bf LHS}}\bra{\Phi^+}\eta_{AA'}\ket{\Phi^+}\le5/8,
}
\end{align}
leading to Eq.~\eqref{Eq:qubit R}.

\section{Proof of Theorem~\ref{Result:conversion}}
We show the following result for general channel resource theories, which includes Theorem~\ref{Result:conversion} as a special case.
Here, the pair $(\mathfrak{F},\mathfrak{O})$ denotes a general channel resource theory, where $\mathfrak{F}$ is the set of all free channels, and $\mathfrak{O}$ is the set of all allowed operations.
Also, \CYnew{we define
\begin{align}
P_\mathfrak{O}(\mathcal{N},\mathcal{K})\coloneqq\max_{\mathbb{F}\in\mathfrak{O}}P[\mathbb{F}(\mathcal{N}),\mathcal{K}].
\end{align}}
\begin{theorem}\label{Result: general DRT conversion}
Suppose $\mathfrak{O}$ is convex, compact, containing the identity map, and closed under function composition (i.e., $\mathbb{F}\circ\mathbb{F}'\in\mathfrak{O}$ $\forall\,\mathbb{F},\mathbb{F}'\in\mathfrak{O}$).
Let $\mathcal{N},\mathcal{M}$ be two channels.
The following three statements are equivalent:
\begin{enumerate}
\item $\mathbb{F}(\mathcal{N}) = \mathcal{M}$ for some $\mathbb{F}\in\mathfrak{O}$.
\item $P_\mathfrak{O}(\mathcal{N},\mathcal{K})\ge P_\mathfrak{O}(\mathcal{M},\mathcal{K})$ for every ${\rm F_1}$ filter $\mathcal{K}$.
\item \CY{$P_\mathfrak{O}(\mathcal{N},\mathcal{K})\ge P_\mathfrak{O}(\mathcal{M},\mathcal{K})$ for every filter $\mathcal{K}$.}
\end{enumerate}
\end{theorem}
Note that setting $\mathfrak{F}={\bf IA}$ and $\mathfrak{O}={\bf AO}$ gives Theorem~\ref{Result:conversion}.

{\em Proof.}
Statement 1 implies statement 2 due to $P_\mathfrak{O}$'s definition and $\mathfrak{O}$'s closedness under function composition.
\CY{Also, statement 2 and statement 3 are equivalent due to Eq.~\eqref{Eq:thx to referee!}.}
It suffices to show statement 2 implies statement 1.
For any given ${\rm F_1}$ filter $\mathcal{K}(\cdot) = K(\cdot)K^\dagger$ in $AA'$ with $0\le K\le\id$, we can write \CYnew{(again, $\Phi^+\coloneqq\proj{\Phi^+}$) 
\begin{align}
P_\mathfrak{O}(\mathcal{N},\mathcal{K})=\max_{\mathbb{F}\in\mathfrak{O}}{\rm tr}\left[K_{AA'}[\mathbb{F}(\mathcal{N})_A\otimes\mathcal{I}_{A'}](\Phi^+_{AA'})\right].
\end{align}}
Then, statement 2 implies that, 
\begin{align}
0&\le P_\mathfrak{O}(\mathcal{N},\mathcal{K}) - P_\mathfrak{O}(\mathcal{M},\mathcal{K})\nonumber\\
&\le\max_{\mathbb{F}\in\mathfrak{O}}{\rm tr}\left[K_{AA'}([\mathbb{F}(\mathcal{N})-\mathcal{M}]_A\otimes\mathcal{I}_{A'})(\Phi^+_{AA'})\right],
\end{align}
where we have used the fact that $\mathfrak{O}$ contains the identity map ``$(\cdot)\mapsto(\cdot)$''.
Since this is true for every ${\rm F_1}$ filter $\mathcal{K}$, we obtain 
\begin{align}
0\le\min_{0\le K\le\id}\max_{\mathbb{F}\in\mathfrak{O}}{\rm tr}\left[K_{AA'}([\mathbb{F}(\mathcal{N})-\mathcal{M}]_A\otimes\mathcal{I}_{A'})(\Phi^+_{AA'})\right].
\end{align}
Now, ${\rm tr}\left[K_{AA'}([\mathbb{F}(\mathcal{N})-\mathcal{M}]_A\otimes\mathcal{I}_{A'})(\Phi^+_{AA'})\right]$ is linear in $K$ when we fix $\mathbb{F}$, and it is continuous, quasi-concave in $\mathbb{F}$ when we fix $K$.
Also, the set $\{K\,|\,0\le K\le\id\}$ is convex and compact, and $\mathfrak{O}$ is convex.
Hence, we can apply Sion's minimax theorem~\cite{Sion1958,Hidetoshi1988} to swap minimisation and maximisation:
\begin{align}
0\le\max_{\mathbb{F}\in\mathfrak{O}}\min_{0\le K\le\id}{\rm tr}\left[K_{AA'}([\mathbb{F}(\mathcal{N})-\mathcal{M}]_A\otimes\mathcal{I}_{A'})(\Phi^+_{AA'})\right].
\end{align}
Since $\mathfrak{O}$ is compact, there exists $\mathbb{F}_*\in\mathfrak{O}$ achieving
\begin{align}
0\le\min_{0\le K\le\id}{\rm tr}\left[K_{AA'}([\mathbb{F}_*(\mathcal{N})-\mathcal{M}]_A\otimes\mathcal{I}_{A'})(\Phi^+_{AA'})\right],
\end{align}
which is equivalent to
\begin{align}
\CYnew{0\le([\mathbb{F}_*(\mathcal{N})-\mathcal{M}]_A\otimes\mathcal{I}_{A'})(\Phi^+_{AA'}).}
\end{align}
Note that $P\ge0$ and ${\rm tr}(P)=0$ imply $P=0$ \CYnew{($P$'s eigenvalues are non-negative and sum to zero).}
Since the right-hand side of the above equation is the difference between two normalised bipartite states, it is trace-less [note that both $\mathbb{F}_*(\mathcal{N})$ and $\mathcal{M}$ are channels].
Hence, we conclude that 
\CYnew{
\begin{align}
[\mathbb{F}_*(\mathcal{N})_A\otimes\mathcal{I}_{A'}](\Phi^+_{AA'})=(\mathcal{M}_A\otimes\mathcal{I}_{A'})(\Phi^+_{AA'}).
\end{align}
}
By Choi-Jamio\l kowski isomorphism~\cite{Choi1975,Jamiolkowski1972}, we obtain 
\begin{align}
\CYnew{
\mathbb{F}_*(\mathcal{N})=\mathcal{M}.
}
\end{align}
Hence, there exists $\mathbb{F}_*\in\mathfrak{O}$ mapping $\mathcal{N}\mapsto\mathcal{M}$.
\hfill$\square$

\bibliography{Ref.bib}

\begin{thebibliography}{93}%
\makeatletter
\providecommand \@ifxundefined [1]{%
 \@ifx{#1\undefined}
}%
\providecommand \@ifnum [1]{%
 \ifnum #1\expandafter \@firstoftwo
 \else \expandafter \@secondoftwo
 \fi
}%
\providecommand \@ifx [1]{%
 \ifx #1\expandafter \@firstoftwo
 \else \expandafter \@secondoftwo
 \fi
}%
\providecommand \natexlab [1]{#1}%
\providecommand \enquote  [1]{``#1''}%
\providecommand \bibnamefont  [1]{#1}%
\providecommand \bibfnamefont [1]{#1}%
\providecommand \citenamefont [1]{#1}%
\providecommand \href@noop [0]{\@secondoftwo}%
\providecommand \href [0]{\begingroup \@sanitize@url \@href}%
\providecommand \@href[1]{\@@startlink{#1}\@@href}%
\providecommand \@@href[1]{\endgroup#1\@@endlink}%
\providecommand \@sanitize@url [0]{\catcode `\\12\catcode `\$12\catcode `\&12\catcode `\#12\catcode `\^12\catcode `\_12\catcode `\%12\relax}%
\providecommand \@@startlink[1]{}%
\providecommand \@@endlink[0]{}%
\providecommand \url  [0]{\begingroup\@sanitize@url \@url }%
\providecommand \@url [1]{\endgroup\@href {#1}{\urlprefix }}%
\providecommand \urlprefix  [0]{URL }%
\providecommand \Eprint [0]{\href }%
\providecommand \doibase [0]{https://doi.org/}%
\providecommand \selectlanguage [0]{\@gobble}%
\providecommand \bibinfo  [0]{\@secondoftwo}%
\providecommand \bibfield  [0]{\@secondoftwo}%
\providecommand \translation [1]{[#1]}%
\providecommand \BibitemOpen [0]{}%
\providecommand \bibitemStop [0]{}%
\providecommand \bibitemNoStop [0]{.\EOS\space}%
\providecommand \EOS [0]{\spacefactor3000\relax}%
\providecommand \BibitemShut  [1]{\csname bibitem#1\endcsname}%
\let\auto@bib@innerbib\@empty
\bibitem [{\citenamefont {Busch}\ \emph {et~al.}(2014)\citenamefont {Busch}, \citenamefont {Lahti},\ and\ \citenamefont {Werner}}]{Busch2014RMP}%
  \BibitemOpen
  \bibfield  {author} {\bibinfo {author} {\bibfnamefont {P.}~\bibnamefont {Busch}}, \bibinfo {author} {\bibfnamefont {P.}~\bibnamefont {Lahti}},\ and\ \bibinfo {author} {\bibfnamefont {R.~F.}\ \bibnamefont {Werner}},\ }\bibfield  {title} {\bibinfo {title} {Colloquium: Quantum root-mean-square error and measurement uncertainty relations},\ }\href {https://doi.org/10.1103/RevModPhys.86.1261} {\bibfield  {journal} {\bibinfo  {journal} {Rev. Mod. Phys.}\ }\textbf {\bibinfo {volume} {86}},\ \bibinfo {pages} {1261} (\bibinfo {year} {2014})}\BibitemShut {NoStop}%
\bibitem [{\citenamefont {G\"uhne}\ \emph {et~al.}(2023)\citenamefont {G\"uhne}, \citenamefont {Haapasalo}, \citenamefont {Kraft}, \citenamefont {Pellonp\"a\"a},\ and\ \citenamefont {Uola}}]{Otfried2023RMP}%
  \BibitemOpen
  \bibfield  {author} {\bibinfo {author} {\bibfnamefont {O.}~\bibnamefont {G\"uhne}}, \bibinfo {author} {\bibfnamefont {E.}~\bibnamefont {Haapasalo}}, \bibinfo {author} {\bibfnamefont {T.}~\bibnamefont {Kraft}}, \bibinfo {author} {\bibfnamefont {J.-P.}\ \bibnamefont {Pellonp\"a\"a}},\ and\ \bibinfo {author} {\bibfnamefont {R.}~\bibnamefont {Uola}},\ }\bibfield  {title} {\bibinfo {title} {Colloquium: Incompatible measurements in quantum information science},\ }\href {https://doi.org/10.1103/RevModPhys.95.011003} {\bibfield  {journal} {\bibinfo  {journal} {Rev. Mod. Phys.}\ }\textbf {\bibinfo {volume} {95}},\ \bibinfo {pages} {011003} (\bibinfo {year} {2023})}\BibitemShut {NoStop}%
\bibitem [{\citenamefont {Bennett}\ and\ \citenamefont {Brassard}(2014)}]{BB84}%
  \BibitemOpen
  \bibfield  {author} {\bibinfo {author} {\bibfnamefont {C.~H.}\ \bibnamefont {Bennett}}\ and\ \bibinfo {author} {\bibfnamefont {G.}~\bibnamefont {Brassard}},\ }\bibfield  {title} {\bibinfo {title} {Quantum cryptography: Public key distribution and coin tossing},\ }\href {https://doi.org/https://doi.org/10.1016/j.tcs.2014.05.025} {\bibfield  {journal} {\bibinfo  {journal} {Theor. Comput. Sci.}\ }\textbf {\bibinfo {volume} {560}},\ \bibinfo {pages} {7} (\bibinfo {year} {2014})}\BibitemShut {NoStop}%
\bibitem [{\citenamefont {Brunner}\ \emph {et~al.}(2014)\citenamefont {Brunner}, \citenamefont {Cavalcanti}, \citenamefont {Pironio}, \citenamefont {Scarani},\ and\ \citenamefont {Wehner}}]{Brunner2014RMP}%
  \BibitemOpen
  \bibfield  {author} {\bibinfo {author} {\bibfnamefont {N.}~\bibnamefont {Brunner}}, \bibinfo {author} {\bibfnamefont {D.}~\bibnamefont {Cavalcanti}}, \bibinfo {author} {\bibfnamefont {S.}~\bibnamefont {Pironio}}, \bibinfo {author} {\bibfnamefont {V.}~\bibnamefont {Scarani}},\ and\ \bibinfo {author} {\bibfnamefont {S.}~\bibnamefont {Wehner}},\ }\bibfield  {title} {\bibinfo {title} {Bell nonlocality},\ }\href {https://doi.org/10.1103/RevModPhys.86.419} {\bibfield  {journal} {\bibinfo  {journal} {Rev. Mod. Phys.}\ }\textbf {\bibinfo {volume} {86}},\ \bibinfo {pages} {419} (\bibinfo {year} {2014})}\BibitemShut {NoStop}%
\bibitem [{\citenamefont {Quintino}\ \emph {et~al.}(2014)\citenamefont {Quintino}, \citenamefont {V\'ertesi},\ and\ \citenamefont {Brunner}}]{Quintino2014PRL}%
  \BibitemOpen
  \bibfield  {author} {\bibinfo {author} {\bibfnamefont {M.~T.}\ \bibnamefont {Quintino}}, \bibinfo {author} {\bibfnamefont {T.}~\bibnamefont {V\'ertesi}},\ and\ \bibinfo {author} {\bibfnamefont {N.}~\bibnamefont {Brunner}},\ }\bibfield  {title} {\bibinfo {title} {Joint measurability, einstein-podolsky-rosen steering, and bell nonlocality},\ }\href {https://doi.org/10.1103/PhysRevLett.113.160402} {\bibfield  {journal} {\bibinfo  {journal} {Phys. Rev. Lett.}\ }\textbf {\bibinfo {volume} {113}},\ \bibinfo {pages} {160402} (\bibinfo {year} {2014})}\BibitemShut {NoStop}%
\bibitem [{\citenamefont {Cavalcanti}\ and\ \citenamefont {Skrzypczyk}(2016{\natexlab{a}})}]{Cavalcanti2016PRA}%
  \BibitemOpen
  \bibfield  {author} {\bibinfo {author} {\bibfnamefont {D.}~\bibnamefont {Cavalcanti}}\ and\ \bibinfo {author} {\bibfnamefont {P.}~\bibnamefont {Skrzypczyk}},\ }\bibfield  {title} {\bibinfo {title} {Quantitative relations between measurement incompatibility, quantum steering, and nonlocality},\ }\href {https://doi.org/10.1103/PhysRevA.93.052112} {\bibfield  {journal} {\bibinfo  {journal} {Phys. Rev. A}\ }\textbf {\bibinfo {volume} {93}},\ \bibinfo {pages} {052112} (\bibinfo {year} {2016}{\natexlab{a}})}\BibitemShut {NoStop}%
\bibitem [{\citenamefont {Uola}\ \emph {et~al.}(2020)\citenamefont {Uola}, \citenamefont {Costa}, \citenamefont {Nguyen},\ and\ \citenamefont {G\"uhne}}]{UolaRMP2020}%
  \BibitemOpen
  \bibfield  {author} {\bibinfo {author} {\bibfnamefont {R.}~\bibnamefont {Uola}}, \bibinfo {author} {\bibfnamefont {A.~C.~S.}\ \bibnamefont {Costa}}, \bibinfo {author} {\bibfnamefont {H.~C.}\ \bibnamefont {Nguyen}},\ and\ \bibinfo {author} {\bibfnamefont {O.}~\bibnamefont {G\"uhne}},\ }\bibfield  {title} {\bibinfo {title} {Quantum steering},\ }\href {https://doi.org/10.1103/RevModPhys.92.015001} {\bibfield  {journal} {\bibinfo  {journal} {Rev. Mod. Phys.}\ }\textbf {\bibinfo {volume} {92}},\ \bibinfo {pages} {015001} (\bibinfo {year} {2020})}\BibitemShut {NoStop}%
\bibitem [{\citenamefont {Cavalcanti}\ and\ \citenamefont {Skrzypczyk}(2016{\natexlab{b}})}]{Cavalcanti2016}%
  \BibitemOpen
  \bibfield  {author} {\bibinfo {author} {\bibfnamefont {D.}~\bibnamefont {Cavalcanti}}\ and\ \bibinfo {author} {\bibfnamefont {P.}~\bibnamefont {Skrzypczyk}},\ }\bibfield  {title} {\bibinfo {title} {Quantum steering: a review with focus on semidefinite programming},\ }\href {https://doi.org/10.1088/1361-6633/80/2/024001} {\bibfield  {journal} {\bibinfo  {journal} {Rep. Prog. Phys.}\ }\textbf {\bibinfo {volume} {80}},\ \bibinfo {pages} {024001} (\bibinfo {year} {2016}{\natexlab{b}})}\BibitemShut {NoStop}%
\bibitem [{\citenamefont {Uola}\ \emph {et~al.}(2014)\citenamefont {Uola}, \citenamefont {Moroder},\ and\ \citenamefont {G\"uhne}}]{Uola2014PRL}%
  \BibitemOpen
  \bibfield  {author} {\bibinfo {author} {\bibfnamefont {R.}~\bibnamefont {Uola}}, \bibinfo {author} {\bibfnamefont {T.}~\bibnamefont {Moroder}},\ and\ \bibinfo {author} {\bibfnamefont {O.}~\bibnamefont {G\"uhne}},\ }\bibfield  {title} {\bibinfo {title} {Joint measurability of generalized measurements implies classicality},\ }\href {https://doi.org/10.1103/PhysRevLett.113.160403} {\bibfield  {journal} {\bibinfo  {journal} {Phys. Rev. Lett.}\ }\textbf {\bibinfo {volume} {113}},\ \bibinfo {pages} {160403} (\bibinfo {year} {2014})}\BibitemShut {NoStop}%
\bibitem [{\citenamefont {Uola}\ \emph {et~al.}(2015)\citenamefont {Uola}, \citenamefont {Budroni}, \citenamefont {G\"uhne},\ and\ \citenamefont {Pellonp\"a\"a}}]{Uola2015PRL}%
  \BibitemOpen
  \bibfield  {author} {\bibinfo {author} {\bibfnamefont {R.}~\bibnamefont {Uola}}, \bibinfo {author} {\bibfnamefont {C.}~\bibnamefont {Budroni}}, \bibinfo {author} {\bibfnamefont {O.}~\bibnamefont {G\"uhne}},\ and\ \bibinfo {author} {\bibfnamefont {J.-P.}\ \bibnamefont {Pellonp\"a\"a}},\ }\bibfield  {title} {\bibinfo {title} {One-to-one mapping between steering and joint measurability problems},\ }\href {https://doi.org/10.1103/PhysRevLett.115.230402} {\bibfield  {journal} {\bibinfo  {journal} {Phys. Rev. Lett.}\ }\textbf {\bibinfo {volume} {115}},\ \bibinfo {pages} {230402} (\bibinfo {year} {2015})}\BibitemShut {NoStop}%
\bibitem [{\citenamefont {Zhao}\ \emph {et~al.}(2020)\citenamefont {Zhao}, \citenamefont {Ku}, \citenamefont {Chen}, \citenamefont {Chen}, \citenamefont {Nori}, \citenamefont {Xiang}, \citenamefont {Li}, \citenamefont {Guo},\ and\ \citenamefont {Chen}}]{Zhao2020}%
  \BibitemOpen
  \bibfield  {author} {\bibinfo {author} {\bibfnamefont {Y.-Y.}\ \bibnamefont {Zhao}}, \bibinfo {author} {\bibfnamefont {H.-Y.}\ \bibnamefont {Ku}}, \bibinfo {author} {\bibfnamefont {S.-L.}\ \bibnamefont {Chen}}, \bibinfo {author} {\bibfnamefont {H.-B.}\ \bibnamefont {Chen}}, \bibinfo {author} {\bibfnamefont {F.}~\bibnamefont {Nori}}, \bibinfo {author} {\bibfnamefont {G.-Y.}\ \bibnamefont {Xiang}}, \bibinfo {author} {\bibfnamefont {C.-F.}\ \bibnamefont {Li}}, \bibinfo {author} {\bibfnamefont {G.-C.}\ \bibnamefont {Guo}},\ and\ \bibinfo {author} {\bibfnamefont {Y.-N.}\ \bibnamefont {Chen}},\ }\bibfield  {title} {\bibinfo {title} {Experimental demonstration of measurement-device-independent measure of quantum steering},\ }\href {https://doi.org/10.1038/s41534-020-00307-9} {\bibfield  {journal} {\bibinfo  {journal} {npj Quantum Inf.}\ }\textbf {\bibinfo {volume} {6}},\ \bibinfo {pages} {77} (\bibinfo {year} {2020})}\BibitemShut {NoStop}%
\bibitem [{\citenamefont {Hsieh}\ \emph {et~al.}()\citenamefont {Hsieh}, \citenamefont {Uola},\ and\ \citenamefont {Skrzypczyk}}]{Hsieh-IP}%
  \BibitemOpen
  \bibfield  {author} {\bibinfo {author} {\bibfnamefont {C.-Y.}\ \bibnamefont {Hsieh}}, \bibinfo {author} {\bibfnamefont {R.}~\bibnamefont {Uola}},\ and\ \bibinfo {author} {\bibfnamefont {P.}~\bibnamefont {Skrzypczyk}},\ }\href@noop {} {\bibinfo {title} {Quantum complementarity: A novel resource for unambiguous exclusion and encryption}},\ \Eprint {https://arxiv.org/abs/2309.11968} {arXiv:2309.11968} \BibitemShut {NoStop}%
\bibitem [{\citenamefont {Ac\'{\i}n}\ \emph {et~al.}(2007)\citenamefont {Ac\'{\i}n}, \citenamefont {Brunner}, \citenamefont {Gisin}, \citenamefont {Massar}, \citenamefont {Pironio},\ and\ \citenamefont {Scarani}}]{Acin2007PRL}%
  \BibitemOpen
  \bibfield  {author} {\bibinfo {author} {\bibfnamefont {A.}~\bibnamefont {Ac\'{\i}n}}, \bibinfo {author} {\bibfnamefont {N.}~\bibnamefont {Brunner}}, \bibinfo {author} {\bibfnamefont {N.}~\bibnamefont {Gisin}}, \bibinfo {author} {\bibfnamefont {S.}~\bibnamefont {Massar}}, \bibinfo {author} {\bibfnamefont {S.}~\bibnamefont {Pironio}},\ and\ \bibinfo {author} {\bibfnamefont {V.}~\bibnamefont {Scarani}},\ }\bibfield  {title} {\bibinfo {title} {Device-independent security of quantum cryptography against collective attacks},\ }\href {https://doi.org/10.1103/PhysRevLett.98.230501} {\bibfield  {journal} {\bibinfo  {journal} {Phys. Rev. Lett.}\ }\textbf {\bibinfo {volume} {98}},\ \bibinfo {pages} {230501} (\bibinfo {year} {2007})}\BibitemShut {NoStop}%
\bibitem [{\citenamefont {Branciard}\ \emph {et~al.}(2012)\citenamefont {Branciard}, \citenamefont {Cavalcanti}, \citenamefont {Walborn}, \citenamefont {Scarani},\ and\ \citenamefont {Wiseman}}]{Branciard2012PRA}%
  \BibitemOpen
  \bibfield  {author} {\bibinfo {author} {\bibfnamefont {C.}~\bibnamefont {Branciard}}, \bibinfo {author} {\bibfnamefont {E.~G.}\ \bibnamefont {Cavalcanti}}, \bibinfo {author} {\bibfnamefont {S.~P.}\ \bibnamefont {Walborn}}, \bibinfo {author} {\bibfnamefont {V.}~\bibnamefont {Scarani}},\ and\ \bibinfo {author} {\bibfnamefont {H.~M.}\ \bibnamefont {Wiseman}},\ }\bibfield  {title} {\bibinfo {title} {One-sided device-independent quantum key distribution: Security, feasibility, and the connection with steering},\ }\href {https://doi.org/10.1103/PhysRevA.85.010301} {\bibfield  {journal} {\bibinfo  {journal} {Phys. Rev. A}\ }\textbf {\bibinfo {volume} {85}},\ \bibinfo {pages} {010301(R)} (\bibinfo {year} {2012})}\BibitemShut {NoStop}%
\bibitem [{\citenamefont {Pironio}\ \emph {et~al.}(2009)\citenamefont {Pironio}, \citenamefont {Ac\'in}, \citenamefont {Brunner}, \citenamefont {Gisin}, \citenamefont {Massar},\ and\ \citenamefont {Scarani}}]{Pironio2009NJP}%
  \BibitemOpen
  \bibfield  {author} {\bibinfo {author} {\bibfnamefont {S.}~\bibnamefont {Pironio}}, \bibinfo {author} {\bibfnamefont {A.}~\bibnamefont {Ac\'in}}, \bibinfo {author} {\bibfnamefont {N.}~\bibnamefont {Brunner}}, \bibinfo {author} {\bibfnamefont {N.}~\bibnamefont {Gisin}}, \bibinfo {author} {\bibfnamefont {S.}~\bibnamefont {Massar}},\ and\ \bibinfo {author} {\bibfnamefont {V.}~\bibnamefont {Scarani}},\ }\bibfield  {title} {\bibinfo {title} {Device-independent quantum key distribution secure against collective attacks},\ }\href {https://doi.org/10.1088/1367-2630/11/4/045021} {\bibfield  {journal} {\bibinfo  {journal} {New J. Phys.}\ }\textbf {\bibinfo {volume} {11}},\ \bibinfo {pages} {045021} (\bibinfo {year} {2009})}\BibitemShut {NoStop}%
\bibitem [{\citenamefont {Ac\'{\i}n}\ \emph {et~al.}(2006)\citenamefont {Ac\'{\i}n}, \citenamefont {Gisin},\ and\ \citenamefont {Masanes}}]{Acin2006PRL}%
  \BibitemOpen
  \bibfield  {author} {\bibinfo {author} {\bibfnamefont {A.}~\bibnamefont {Ac\'{\i}n}}, \bibinfo {author} {\bibfnamefont {N.}~\bibnamefont {Gisin}},\ and\ \bibinfo {author} {\bibfnamefont {L.}~\bibnamefont {Masanes}},\ }\bibfield  {title} {\bibinfo {title} {From bell's theorem to secure quantum key distribution},\ }\href {https://doi.org/10.1103/PhysRevLett.97.120405} {\bibfield  {journal} {\bibinfo  {journal} {Phys. Rev. Lett.}\ }\textbf {\bibinfo {volume} {97}},\ \bibinfo {pages} {120405} (\bibinfo {year} {2006})}\BibitemShut {NoStop}%
\bibitem [{\citenamefont {Buscemi}\ \emph {et~al.}(2023)\citenamefont {Buscemi}, \citenamefont {Kobayashi}, \citenamefont {Minagawa}, \citenamefont {Perinotti},\ and\ \citenamefont {Tosini}}]{Buscemi2023Quantum}%
  \BibitemOpen
  \bibfield  {author} {\bibinfo {author} {\bibfnamefont {F.}~\bibnamefont {Buscemi}}, \bibinfo {author} {\bibfnamefont {K.}~\bibnamefont {Kobayashi}}, \bibinfo {author} {\bibfnamefont {S.}~\bibnamefont {Minagawa}}, \bibinfo {author} {\bibfnamefont {P.}~\bibnamefont {Perinotti}},\ and\ \bibinfo {author} {\bibfnamefont {A.}~\bibnamefont {Tosini}},\ }\bibfield  {title} {\bibinfo {title} {Unifying different notions of quantum incompatibility into a strict hierarchy of resource theories of communication},\ }\href {https://doi.org/10.22331/q-2023-06-07-1035} {\bibfield  {journal} {\bibinfo  {journal} {{Quantum}}\ }\textbf {\bibinfo {volume} {7}},\ \bibinfo {pages} {1035} (\bibinfo {year} {2023})}\BibitemShut {NoStop}%
\bibitem [{\citenamefont {Buscemi}\ \emph {et~al.}(2020)\citenamefont {Buscemi}, \citenamefont {Chitambar},\ and\ \citenamefont {Zhou}}]{Buscemi2020PRL}%
  \BibitemOpen
  \bibfield  {author} {\bibinfo {author} {\bibfnamefont {F.}~\bibnamefont {Buscemi}}, \bibinfo {author} {\bibfnamefont {E.}~\bibnamefont {Chitambar}},\ and\ \bibinfo {author} {\bibfnamefont {W.}~\bibnamefont {Zhou}},\ }\bibfield  {title} {\bibinfo {title} {Complete resource theory of quantum incompatibility as quantum programmability},\ }\href {https://doi.org/10.1103/PhysRevLett.124.120401} {\bibfield  {journal} {\bibinfo  {journal} {Phys. Rev. Lett.}\ }\textbf {\bibinfo {volume} {124}},\ \bibinfo {pages} {120401} (\bibinfo {year} {2020})}\BibitemShut {NoStop}%
\bibitem [{\citenamefont {Ji}\ and\ \citenamefont {Chitambar}(2024)}]{Ji2024PRXQ}%
  \BibitemOpen
  \bibfield  {author} {\bibinfo {author} {\bibfnamefont {K.}~\bibnamefont {Ji}}\ and\ \bibinfo {author} {\bibfnamefont {E.}~\bibnamefont {Chitambar}},\ }\bibfield  {title} {\bibinfo {title} {Incompatibility as a resource for programmable quantum instruments},\ }\href {https://doi.org/10.1103/PRXQuantum.5.010340} {\bibfield  {journal} {\bibinfo  {journal} {PRX Quantum}\ }\textbf {\bibinfo {volume} {5}},\ \bibinfo {pages} {010340} (\bibinfo {year} {2024})}\BibitemShut {NoStop}%
\bibitem [{\citenamefont {Skrzypczyk}\ \emph {et~al.}(2019)\citenamefont {Skrzypczyk}, \citenamefont {\ifmmode \check{S}\else \v{S}\fi{}upi\ifmmode~\acute{c}\else \'{c}\fi{}},\ and\ \citenamefont {Cavalcanti}}]{Skrzypczyk2019PRL}%
  \BibitemOpen
  \bibfield  {author} {\bibinfo {author} {\bibfnamefont {P.}~\bibnamefont {Skrzypczyk}}, \bibinfo {author} {\bibfnamefont {I.}~\bibnamefont {\ifmmode \check{S}\else \v{S}\fi{}upi\ifmmode~\acute{c}\else \'{c}\fi{}}},\ and\ \bibinfo {author} {\bibfnamefont {D.}~\bibnamefont {Cavalcanti}},\ }\bibfield  {title} {\bibinfo {title} {All sets of incompatible measurements give an advantage in quantum state discrimination},\ }\href {https://doi.org/10.1103/PhysRevLett.122.130403} {\bibfield  {journal} {\bibinfo  {journal} {Phys. Rev. Lett.}\ }\textbf {\bibinfo {volume} {122}},\ \bibinfo {pages} {130403} (\bibinfo {year} {2019})}\BibitemShut {NoStop}%
\bibitem [{\citenamefont {Ku}\ \emph {et~al.}(2022{\natexlab{a}})\citenamefont {Ku}, \citenamefont {Hsieh}, \citenamefont {Chen}, \citenamefont {Chen},\ and\ \citenamefont {Budroni}}]{Ku2022NC}%
  \BibitemOpen
  \bibfield  {author} {\bibinfo {author} {\bibfnamefont {H.-Y.}\ \bibnamefont {Ku}}, \bibinfo {author} {\bibfnamefont {C.-Y.}\ \bibnamefont {Hsieh}}, \bibinfo {author} {\bibfnamefont {S.-L.}\ \bibnamefont {Chen}}, \bibinfo {author} {\bibfnamefont {Y.-N.}\ \bibnamefont {Chen}},\ and\ \bibinfo {author} {\bibfnamefont {C.}~\bibnamefont {Budroni}},\ }\bibfield  {title} {\bibinfo {title} {Complete classification of steerability under local filters and its relation with measurement incompatibility},\ }\href {https://doi.org/10.1038/s41467-022-32466-y} {\bibfield  {journal} {\bibinfo  {journal} {Nat. Commun.}\ }\textbf {\bibinfo {volume} {13}},\ \bibinfo {pages} {4973} (\bibinfo {year} {2022}{\natexlab{a}})}\BibitemShut {NoStop}%
\bibitem [{\citenamefont {Ku}\ \emph {et~al.}()\citenamefont {Ku}, \citenamefont {Hsieh},\ and\ \citenamefont {Budroni}}]{ku2023}%
  \BibitemOpen
  \bibfield  {author} {\bibinfo {author} {\bibfnamefont {H.-Y.}\ \bibnamefont {Ku}}, \bibinfo {author} {\bibfnamefont {C.-Y.}\ \bibnamefont {Hsieh}},\ and\ \bibinfo {author} {\bibfnamefont {C.}~\bibnamefont {Budroni}},\ }\href@noop {} {\bibinfo {title} {Measurement incompatibility cannot be stochastically distilled}},\ \Eprint {https://arxiv.org/abs/2308.02252} {arXiv:2308.02252} \BibitemShut {NoStop}%
\bibitem [{\citenamefont {Hsieh}\ \emph {et~al.}(2023)\citenamefont {Hsieh}, \citenamefont {Ku},\ and\ \citenamefont {Budroni}}]{Hsieh2023}%
  \BibitemOpen
  \bibfield  {author} {\bibinfo {author} {\bibfnamefont {C.-Y.}\ \bibnamefont {Hsieh}}, \bibinfo {author} {\bibfnamefont {H.-Y.}\ \bibnamefont {Ku}},\ and\ \bibinfo {author} {\bibfnamefont {C.}~\bibnamefont {Budroni}},\ }\href@noop {} {\bibinfo {title} {Characterisation and fundamental limitations of irreversible stochastic steering distillation}} (\bibinfo {year} {2023}),\ \Eprint {https://arxiv.org/abs/2309.06191} {arXiv:2309.06191} \BibitemShut {NoStop}%
\bibitem [{\citenamefont {Pironio}\ \emph {et~al.}(2010)\citenamefont {Pironio}, \citenamefont {Ac{\'i}n}, \citenamefont {Massar}, \citenamefont {de~la Giroday}, \citenamefont {Matsukevich}, \citenamefont {Maunz}, \citenamefont {Olmschenk}, \citenamefont {Hayes}, \citenamefont {Luo}, \citenamefont {Manning},\ and\ \citenamefont {Monroe}}]{Pironio2010Nature}%
  \BibitemOpen
  \bibfield  {author} {\bibinfo {author} {\bibfnamefont {S.}~\bibnamefont {Pironio}}, \bibinfo {author} {\bibfnamefont {A.}~\bibnamefont {Ac{\'i}n}}, \bibinfo {author} {\bibfnamefont {S.}~\bibnamefont {Massar}}, \bibinfo {author} {\bibfnamefont {A.~B.}\ \bibnamefont {de~la Giroday}}, \bibinfo {author} {\bibfnamefont {D.~N.}\ \bibnamefont {Matsukevich}}, \bibinfo {author} {\bibfnamefont {P.}~\bibnamefont {Maunz}}, \bibinfo {author} {\bibfnamefont {S.}~\bibnamefont {Olmschenk}}, \bibinfo {author} {\bibfnamefont {D.}~\bibnamefont {Hayes}}, \bibinfo {author} {\bibfnamefont {L.}~\bibnamefont {Luo}}, \bibinfo {author} {\bibfnamefont {T.~A.}\ \bibnamefont {Manning}},\ and\ \bibinfo {author} {\bibfnamefont {C.}~\bibnamefont {Monroe}},\ }\bibfield  {title} {\bibinfo {title} {Random numbers certified by {B}ell's theorem},\ }\href {https://doi.org/10.1038/nature09008} {\bibfield  {journal} {\bibinfo  {journal} {Nature}\ }\textbf {\bibinfo {volume} {464}},\ \bibinfo {pages} {1021} (\bibinfo {year} {2010})}\BibitemShut
  {NoStop}%
\bibitem [{\citenamefont {Skrzypczyk}\ and\ \citenamefont {Cavalcanti}(2018)}]{SkrzypczykPRL2018}%
  \BibitemOpen
  \bibfield  {author} {\bibinfo {author} {\bibfnamefont {P.}~\bibnamefont {Skrzypczyk}}\ and\ \bibinfo {author} {\bibfnamefont {D.}~\bibnamefont {Cavalcanti}},\ }\bibfield  {title} {\bibinfo {title} {Maximal randomness generation from steering inequality violations using qudits},\ }\href {https://doi.org/10.1103/PhysRevLett.120.260401} {\bibfield  {journal} {\bibinfo  {journal} {Phys. Rev. Lett.}\ }\textbf {\bibinfo {volume} {120}},\ \bibinfo {pages} {260401} (\bibinfo {year} {2018})}\BibitemShut {NoStop}%
\bibitem [{\citenamefont {Hsieh}\ and\ \citenamefont {Chen}(2024)}]{Hsieh2024}%
  \BibitemOpen
  \bibfield  {author} {\bibinfo {author} {\bibfnamefont {C.-Y.}\ \bibnamefont {Hsieh}}\ and\ \bibinfo {author} {\bibfnamefont {S.-L.}\ \bibnamefont {Chen}},\ }\bibfield  {title} {\bibinfo {title} {Thermodynamic approach to quantifying incompatible instruments},\ }\href {https://doi.org/10.1103/PhysRevLett.133.170401} {\bibfield  {journal} {\bibinfo  {journal} {Phys. Rev. Lett.}\ }\textbf {\bibinfo {volume} {133}},\ \bibinfo {pages} {170401} (\bibinfo {year} {2024})}\BibitemShut {NoStop}%
\bibitem [{\citenamefont {Hsieh}\ and\ \citenamefont {Gessner}(2024)}]{Hsieh2023IP}%
  \BibitemOpen
  \bibfield  {author} {\bibinfo {author} {\bibfnamefont {C.-Y.}\ \bibnamefont {Hsieh}}\ and\ \bibinfo {author} {\bibfnamefont {M.}~\bibnamefont {Gessner}},\ }\href@noop {} {\bibinfo {title} {General quantum resources provide advantages in work extraction tasks}} (\bibinfo {year} {2024}),\ \Eprint {https://arxiv.org/abs/2403.18753} {arXiv:2403.18753} \BibitemShut {NoStop}%
\bibitem [{\citenamefont {Kosloff}\ and\ \citenamefont {Feldmann}(2002)}]{KosloffPRE2002}%
  \BibitemOpen
  \bibfield  {author} {\bibinfo {author} {\bibfnamefont {R.}~\bibnamefont {Kosloff}}\ and\ \bibinfo {author} {\bibfnamefont {T.}~\bibnamefont {Feldmann}},\ }\bibfield  {title} {\bibinfo {title} {Discrete four-stroke quantum heat engine exploring the origin of friction},\ }\href {https://doi.org/10.1103/PhysRevE.65.055102} {\bibfield  {journal} {\bibinfo  {journal} {Phys. Rev. E}\ }\textbf {\bibinfo {volume} {65}},\ \bibinfo {pages} {055102} (\bibinfo {year} {2002})}\BibitemShut {NoStop}%
\bibitem [{\citenamefont {Feldmann}\ and\ \citenamefont {Kosloff}(2006)}]{FeldmannPRE2006}%
  \BibitemOpen
  \bibfield  {author} {\bibinfo {author} {\bibfnamefont {T.}~\bibnamefont {Feldmann}}\ and\ \bibinfo {author} {\bibfnamefont {R.}~\bibnamefont {Kosloff}},\ }\bibfield  {title} {\bibinfo {title} {Quantum lubrication: Suppression of friction in a first-principles four-stroke heat engine},\ }\href {https://doi.org/10.1103/PhysRevE.73.025107} {\bibfield  {journal} {\bibinfo  {journal} {Phys. Rev. E}\ }\textbf {\bibinfo {volume} {73}},\ \bibinfo {pages} {025107} (\bibinfo {year} {2006})}\BibitemShut {NoStop}%
\bibitem [{\citenamefont {Lostaglio}\ \emph {et~al.}(2017)\citenamefont {Lostaglio}, \citenamefont {Jennings},\ and\ \citenamefont {Rudolph}}]{LostaglioNJP2017}%
  \BibitemOpen
  \bibfield  {author} {\bibinfo {author} {\bibfnamefont {M.}~\bibnamefont {Lostaglio}}, \bibinfo {author} {\bibfnamefont {D.}~\bibnamefont {Jennings}},\ and\ \bibinfo {author} {\bibfnamefont {T.}~\bibnamefont {Rudolph}},\ }\bibfield  {title} {\bibinfo {title} {Thermodynamic resource theories, non-commutativity and maximum entropy principles},\ }\href {https://doi.org/10.1088/1367-2630/aa617f} {\bibfield  {journal} {\bibinfo  {journal} {New J. Phys.}\ }\textbf {\bibinfo {volume} {19}},\ \bibinfo {pages} {043008} (\bibinfo {year} {2017})}\BibitemShut {NoStop}%
\bibitem [{\citenamefont {Majidy}\ \emph {et~al.}(2023)\citenamefont {Majidy}, \citenamefont {Braasch}, \citenamefont {Lasek}, \citenamefont {Upadhyaya}, \citenamefont {Kalev},\ and\ \citenamefont {Yunger~Halpern}}]{Majidy2023}%
  \BibitemOpen
  \bibfield  {author} {\bibinfo {author} {\bibfnamefont {S.}~\bibnamefont {Majidy}}, \bibinfo {author} {\bibfnamefont {W.~F.}\ \bibnamefont {Braasch}}, \bibinfo {author} {\bibfnamefont {A.}~\bibnamefont {Lasek}}, \bibinfo {author} {\bibfnamefont {T.}~\bibnamefont {Upadhyaya}}, \bibinfo {author} {\bibfnamefont {A.}~\bibnamefont {Kalev}},\ and\ \bibinfo {author} {\bibfnamefont {N.}~\bibnamefont {Yunger~Halpern}},\ }\bibfield  {title} {\bibinfo {title} {Noncommuting conserved charges in quantum thermodynamics and beyond},\ }\href {https://doi.org/10.1038/s42254-023-00641-9} {\bibfield  {journal} {\bibinfo  {journal} {Nat. Rev. Phys.}\ }\textbf {\bibinfo {volume} {5}},\ \bibinfo {pages} {689} (\bibinfo {year} {2023})}\BibitemShut {NoStop}%
\bibitem [{\citenamefont {Yunger~Halpern}\ \emph {et~al.}(2016)\citenamefont {Yunger~Halpern}, \citenamefont {Faist}, \citenamefont {Oppenheim},\ and\ \citenamefont {Winter}}]{YungerHalpern2016NC}%
  \BibitemOpen
  \bibfield  {author} {\bibinfo {author} {\bibfnamefont {N.}~\bibnamefont {Yunger~Halpern}}, \bibinfo {author} {\bibfnamefont {P.}~\bibnamefont {Faist}}, \bibinfo {author} {\bibfnamefont {J.}~\bibnamefont {Oppenheim}},\ and\ \bibinfo {author} {\bibfnamefont {A.}~\bibnamefont {Winter}},\ }\bibfield  {title} {\bibinfo {title} {Microcanonical and resource-theoretic derivations of the thermal state of a quantum system with noncommuting charges},\ }\href {https://doi.org/10.1038/ncomms12051} {\bibfield  {journal} {\bibinfo  {journal} {Nat. Commun.}\ }\textbf {\bibinfo {volume} {7}},\ \bibinfo {pages} {12051} (\bibinfo {year} {2016})}\BibitemShut {NoStop}%
\bibitem [{\citenamefont {Guryanova}\ \emph {et~al.}(2016)\citenamefont {Guryanova}, \citenamefont {Popescu}, \citenamefont {Short}, \citenamefont {Silva},\ and\ \citenamefont {Skrzypczyk}}]{Guryanova2016NC}%
  \BibitemOpen
  \bibfield  {author} {\bibinfo {author} {\bibfnamefont {Y.}~\bibnamefont {Guryanova}}, \bibinfo {author} {\bibfnamefont {S.}~\bibnamefont {Popescu}}, \bibinfo {author} {\bibfnamefont {A.~J.}\ \bibnamefont {Short}}, \bibinfo {author} {\bibfnamefont {R.}~\bibnamefont {Silva}},\ and\ \bibinfo {author} {\bibfnamefont {P.}~\bibnamefont {Skrzypczyk}},\ }\bibfield  {title} {\bibinfo {title} {Thermodynamics of quantum systems with multiple conserved quantities},\ }\href {https://doi.org/10.1038/ncomms12049} {\bibfield  {journal} {\bibinfo  {journal} {Nat. Commun.}\ }\textbf {\bibinfo {volume} {7}},\ \bibinfo {pages} {12049} (\bibinfo {year} {2016})}\BibitemShut {NoStop}%
\bibitem [{\citenamefont {Ji}\ \emph {et~al.}(2022)\citenamefont {Ji}, \citenamefont {Chai}, \citenamefont {Wang}, \citenamefont {Guo}, \citenamefont {Rong}, \citenamefont {Shi}, \citenamefont {Ren}, \citenamefont {Wang},\ and\ \citenamefont {Du}}]{JiPRL2022}%
  \BibitemOpen
  \bibfield  {author} {\bibinfo {author} {\bibfnamefont {W.}~\bibnamefont {Ji}}, \bibinfo {author} {\bibfnamefont {Z.}~\bibnamefont {Chai}}, \bibinfo {author} {\bibfnamefont {M.}~\bibnamefont {Wang}}, \bibinfo {author} {\bibfnamefont {Y.}~\bibnamefont {Guo}}, \bibinfo {author} {\bibfnamefont {X.}~\bibnamefont {Rong}}, \bibinfo {author} {\bibfnamefont {F.}~\bibnamefont {Shi}}, \bibinfo {author} {\bibfnamefont {C.}~\bibnamefont {Ren}}, \bibinfo {author} {\bibfnamefont {Y.}~\bibnamefont {Wang}},\ and\ \bibinfo {author} {\bibfnamefont {J.}~\bibnamefont {Du}},\ }\bibfield  {title} {\bibinfo {title} {Spin quantum heat engine quantified by quantum steering},\ }\href {https://doi.org/10.1103/PhysRevLett.128.090602} {\bibfield  {journal} {\bibinfo  {journal} {Phys. Rev. Lett.}\ }\textbf {\bibinfo {volume} {128}},\ \bibinfo {pages} {090602} (\bibinfo {year} {2022})}\BibitemShut {NoStop}%
\bibitem [{\citenamefont {Beyer}\ \emph {et~al.}(2019)\citenamefont {Beyer}, \citenamefont {Luoma},\ and\ \citenamefont {Strunz}}]{BeyerPRL2019}%
  \BibitemOpen
  \bibfield  {author} {\bibinfo {author} {\bibfnamefont {K.}~\bibnamefont {Beyer}}, \bibinfo {author} {\bibfnamefont {K.}~\bibnamefont {Luoma}},\ and\ \bibinfo {author} {\bibfnamefont {W.~T.}\ \bibnamefont {Strunz}},\ }\bibfield  {title} {\bibinfo {title} {Steering heat engines: A truly quantum maxwell demon},\ }\href {https://doi.org/10.1103/PhysRevLett.123.250606} {\bibfield  {journal} {\bibinfo  {journal} {Phys. Rev. Lett.}\ }\textbf {\bibinfo {volume} {123}},\ \bibinfo {pages} {250606} (\bibinfo {year} {2019})}\BibitemShut {NoStop}%
\bibitem [{\citenamefont {Chan}\ \emph {et~al.}(2022)\citenamefont {Chan}, \citenamefont {Huang}, \citenamefont {Lin}, \citenamefont {Ku}, \citenamefont {Chen}, \citenamefont {Chen},\ and\ \citenamefont {Chen}}]{ChanPRA2022}%
  \BibitemOpen
  \bibfield  {author} {\bibinfo {author} {\bibfnamefont {F.-J.}\ \bibnamefont {Chan}}, \bibinfo {author} {\bibfnamefont {Y.-T.}\ \bibnamefont {Huang}}, \bibinfo {author} {\bibfnamefont {J.-D.}\ \bibnamefont {Lin}}, \bibinfo {author} {\bibfnamefont {H.-Y.}\ \bibnamefont {Ku}}, \bibinfo {author} {\bibfnamefont {J.-S.}\ \bibnamefont {Chen}}, \bibinfo {author} {\bibfnamefont {H.-B.}\ \bibnamefont {Chen}},\ and\ \bibinfo {author} {\bibfnamefont {Y.-N.}\ \bibnamefont {Chen}},\ }\bibfield  {title} {\bibinfo {title} {Maxwell's two-demon engine under pure dephasing noise},\ }\href {https://doi.org/10.1103/PhysRevA.106.052201} {\bibfield  {journal} {\bibinfo  {journal} {Phys. Rev. A}\ }\textbf {\bibinfo {volume} {106}},\ \bibinfo {pages} {052201} (\bibinfo {year} {2022})}\BibitemShut {NoStop}%
\bibitem [{\citenamefont {Wilde}(2017)}]{wilde_2017}%
  \BibitemOpen
  \bibfield  {author} {\bibinfo {author} {\bibfnamefont {M.~M.}\ \bibnamefont {Wilde}},\ }\href {https://doi.org/10.1017/9781316809976} {\emph {\bibinfo {title} {Quantum Information Theory}}},\ \bibinfo {edition} {2nd}\ ed.\ (\bibinfo  {publisher} {Cambridge University Press},\ \bibinfo {year} {2017})\BibitemShut {NoStop}%
\bibitem [{\citenamefont {Rosset}\ \emph {et~al.}(2018)\citenamefont {Rosset}, \citenamefont {Buscemi},\ and\ \citenamefont {Liang}}]{RossetPRX2018}%
  \BibitemOpen
  \bibfield  {author} {\bibinfo {author} {\bibfnamefont {D.}~\bibnamefont {Rosset}}, \bibinfo {author} {\bibfnamefont {F.}~\bibnamefont {Buscemi}},\ and\ \bibinfo {author} {\bibfnamefont {Y.-C.}\ \bibnamefont {Liang}},\ }\bibfield  {title} {\bibinfo {title} {Resource theory of quantum memories and their faithful verification with minimal assumptions},\ }\href {https://doi.org/10.1103/PhysRevX.8.021033} {\bibfield  {journal} {\bibinfo  {journal} {Phys. Rev. X}\ }\textbf {\bibinfo {volume} {8}},\ \bibinfo {pages} {021033} (\bibinfo {year} {2018})}\BibitemShut {NoStop}%
\bibitem [{\citenamefont {Ku}\ \emph {et~al.}(2022{\natexlab{b}})\citenamefont {Ku}, \citenamefont {Kadlec}, \citenamefont {\ifmmode~\check{C}\else \v{C}\fi{}ernoch}, \citenamefont {Quintino}, \citenamefont {Zhou}, \citenamefont {Lemr}, \citenamefont {Lambert}, \citenamefont {Miranowicz}, \citenamefont {Chen}, \citenamefont {Nori},\ and\ \citenamefont {Chen}}]{Ku2022PRXQ}%
  \BibitemOpen
  \bibfield  {author} {\bibinfo {author} {\bibfnamefont {H.-Y.}\ \bibnamefont {Ku}}, \bibinfo {author} {\bibfnamefont {J.}~\bibnamefont {Kadlec}}, \bibinfo {author} {\bibfnamefont {A.}~\bibnamefont {\ifmmode~\check{C}\else \v{C}\fi{}ernoch}}, \bibinfo {author} {\bibfnamefont {M.~T.}\ \bibnamefont {Quintino}}, \bibinfo {author} {\bibfnamefont {W.}~\bibnamefont {Zhou}}, \bibinfo {author} {\bibfnamefont {K.}~\bibnamefont {Lemr}}, \bibinfo {author} {\bibfnamefont {N.}~\bibnamefont {Lambert}}, \bibinfo {author} {\bibfnamefont {A.}~\bibnamefont {Miranowicz}}, \bibinfo {author} {\bibfnamefont {S.-L.}\ \bibnamefont {Chen}}, \bibinfo {author} {\bibfnamefont {F.}~\bibnamefont {Nori}},\ and\ \bibinfo {author} {\bibfnamefont {Y.-N.}\ \bibnamefont {Chen}},\ }\bibfield  {title} {\bibinfo {title} {Quantifying quantumness of channels without entanglement},\ }\href {https://doi.org/10.1103/PRXQuantum.3.020338} {\bibfield  {journal} {\bibinfo  {journal} {PRX Quantum}\ }\textbf {\bibinfo {volume} {3}},\ \bibinfo {pages}
  {020338} (\bibinfo {year} {2022}{\natexlab{b}})}\BibitemShut {NoStop}%
\bibitem [{\citenamefont {Yuan}\ \emph {et~al.}(2021)\citenamefont {Yuan}, \citenamefont {Liu}, \citenamefont {Zhao}, \citenamefont {Regula}, \citenamefont {Thompson},\ and\ \citenamefont {Gu}}]{Yuan2021npjQI}%
  \BibitemOpen
  \bibfield  {author} {\bibinfo {author} {\bibfnamefont {X.}~\bibnamefont {Yuan}}, \bibinfo {author} {\bibfnamefont {Y.}~\bibnamefont {Liu}}, \bibinfo {author} {\bibfnamefont {Q.}~\bibnamefont {Zhao}}, \bibinfo {author} {\bibfnamefont {B.}~\bibnamefont {Regula}}, \bibinfo {author} {\bibfnamefont {J.}~\bibnamefont {Thompson}},\ and\ \bibinfo {author} {\bibfnamefont {M.}~\bibnamefont {Gu}},\ }\bibfield  {title} {\bibinfo {title} {Universal and operational benchmarking of quantum memories},\ }\href {https://doi.org/10.1038/s41534-021-00444-9} {\bibfield  {journal} {\bibinfo  {journal} {npj Quantum Inf.}\ }\textbf {\bibinfo {volume} {7}},\ \bibinfo {pages} {108} (\bibinfo {year} {2021})}\BibitemShut {NoStop}%
\bibitem [{\citenamefont {Vieira}\ \emph {et~al.}(2024)\citenamefont {Vieira}, \citenamefont {Ku},\ and\ \citenamefont {Budroni}}]{Vieira2024}%
  \BibitemOpen
  \bibfield  {author} {\bibinfo {author} {\bibfnamefont {L.~B.}\ \bibnamefont {Vieira}}, \bibinfo {author} {\bibfnamefont {H.-Y.}\ \bibnamefont {Ku}},\ and\ \bibinfo {author} {\bibfnamefont {C.}~\bibnamefont {Budroni}},\ }\href@noop {} {\bibinfo {title} {Entanglement-breaking channels are a quantum memory resource}} (\bibinfo {year} {2024}),\ \Eprint {https://arxiv.org/abs/2402.16789} {arXiv:2402.16789 [quant-ph]} \BibitemShut {NoStop}%
\bibitem [{\citenamefont {Chang}\ \emph {et~al.}(2024)\citenamefont {Chang}, \citenamefont {Ju}, \citenamefont {Chen}, \citenamefont {Chen},\ and\ \citenamefont {Ku}}]{Chang2024PRR}%
  \BibitemOpen
  \bibfield  {author} {\bibinfo {author} {\bibfnamefont {W.-G.}\ \bibnamefont {Chang}}, \bibinfo {author} {\bibfnamefont {C.-Y.}\ \bibnamefont {Ju}}, \bibinfo {author} {\bibfnamefont {G.-Y.}\ \bibnamefont {Chen}}, \bibinfo {author} {\bibfnamefont {Y.-N.}\ \bibnamefont {Chen}},\ and\ \bibinfo {author} {\bibfnamefont {H.-Y.}\ \bibnamefont {Ku}},\ }\bibfield  {title} {\bibinfo {title} {Visually quantifying single-qubit quantum memory},\ }\href {https://doi.org/10.1103/PhysRevResearch.6.023035} {\bibfield  {journal} {\bibinfo  {journal} {Phys. Rev. Res.}\ }\textbf {\bibinfo {volume} {6}},\ \bibinfo {pages} {023035} (\bibinfo {year} {2024})}\BibitemShut {NoStop}%
\bibitem [{\citenamefont {Hsieh}(2020)}]{Hsieh2020Quantum}%
  \BibitemOpen
  \bibfield  {author} {\bibinfo {author} {\bibfnamefont {C.-Y.}\ \bibnamefont {Hsieh}},\ }\bibfield  {title} {\bibinfo {title} {Resource preservability},\ }\href {https://doi.org/10.22331/q-2020-03-19-244} {\bibfield  {journal} {\bibinfo  {journal} {{Quantum}}\ }\textbf {\bibinfo {volume} {4}},\ \bibinfo {pages} {244} (\bibinfo {year} {2020})}\BibitemShut {NoStop}%
\bibitem [{\citenamefont {Hsieh}(2021)}]{Hsieh2021PRXQ}%
  \BibitemOpen
  \bibfield  {author} {\bibinfo {author} {\bibfnamefont {C.-Y.}\ \bibnamefont {Hsieh}},\ }\bibfield  {title} {\bibinfo {title} {Communication, dynamical resource theory, and thermodynamics},\ }\href {https://doi.org/10.1103/PRXQuantum.2.020318} {\bibfield  {journal} {\bibinfo  {journal} {PRX Quantum}\ }\textbf {\bibinfo {volume} {2}},\ \bibinfo {pages} {020318} (\bibinfo {year} {2021})}\BibitemShut {NoStop}%
\bibitem [{\citenamefont {Stratton}\ \emph {et~al.}(2024{\natexlab{a}})\citenamefont {Stratton}, \citenamefont {Hsieh},\ and\ \citenamefont {Skrzypczyk}}]{Stratton2024PRL}%
  \BibitemOpen
  \bibfield  {author} {\bibinfo {author} {\bibfnamefont {B.}~\bibnamefont {Stratton}}, \bibinfo {author} {\bibfnamefont {C.-Y.}\ \bibnamefont {Hsieh}},\ and\ \bibinfo {author} {\bibfnamefont {P.}~\bibnamefont {Skrzypczyk}},\ }\bibfield  {title} {\bibinfo {title} {Dynamical resource theory of informational nonequilibrium preservability},\ }\href {https://doi.org/10.1103/PhysRevLett.132.110202} {\bibfield  {journal} {\bibinfo  {journal} {Phys. Rev. Lett.}\ }\textbf {\bibinfo {volume} {132}},\ \bibinfo {pages} {110202} (\bibinfo {year} {2024}{\natexlab{a}})}\BibitemShut {NoStop}%
\bibitem [{\citenamefont {Saxena}\ \emph {et~al.}(2020)\citenamefont {Saxena}, \citenamefont {Chitambar},\ and\ \citenamefont {Gour}}]{SaxenaPRR2020}%
  \BibitemOpen
  \bibfield  {author} {\bibinfo {author} {\bibfnamefont {G.}~\bibnamefont {Saxena}}, \bibinfo {author} {\bibfnamefont {E.}~\bibnamefont {Chitambar}},\ and\ \bibinfo {author} {\bibfnamefont {G.}~\bibnamefont {Gour}},\ }\bibfield  {title} {\bibinfo {title} {Dynamical resource theory of quantum coherence},\ }\href {https://doi.org/10.1103/PhysRevResearch.2.023298} {\bibfield  {journal} {\bibinfo  {journal} {Phys. Rev. Res.}\ }\textbf {\bibinfo {volume} {2}},\ \bibinfo {pages} {023298} (\bibinfo {year} {2020})}\BibitemShut {NoStop}%
\bibitem [{\citenamefont {Hsieh}\ \emph {et~al.}(2020)\citenamefont {Hsieh}, \citenamefont {Lostaglio},\ and\ \citenamefont {Ac\'{\i}n}}]{Hsieh2020PRR}%
  \BibitemOpen
  \bibfield  {author} {\bibinfo {author} {\bibfnamefont {C.-Y.}\ \bibnamefont {Hsieh}}, \bibinfo {author} {\bibfnamefont {M.}~\bibnamefont {Lostaglio}},\ and\ \bibinfo {author} {\bibfnamefont {A.}~\bibnamefont {Ac\'{\i}n}},\ }\bibfield  {title} {\bibinfo {title} {Entanglement preserving local thermalization},\ }\href {https://doi.org/10.1103/PhysRevResearch.2.013379} {\bibfield  {journal} {\bibinfo  {journal} {Phys. Rev. Res.}\ }\textbf {\bibinfo {volume} {2}},\ \bibinfo {pages} {013379} (\bibinfo {year} {2020})}\BibitemShut {NoStop}%
\bibitem [{\citenamefont {Chen}\ \emph {et~al.}(2021)\citenamefont {Chen}, \citenamefont {Ng},\ and\ \citenamefont {Li}}]{ChenPRA2021}%
  \BibitemOpen
  \bibfield  {author} {\bibinfo {author} {\bibfnamefont {S.-H.}\ \bibnamefont {Chen}}, \bibinfo {author} {\bibfnamefont {M.-L.}\ \bibnamefont {Ng}},\ and\ \bibinfo {author} {\bibfnamefont {C.-M.}\ \bibnamefont {Li}},\ }\bibfield  {title} {\bibinfo {title} {Quantifying entanglement preservability of experimental processes},\ }\href {https://doi.org/10.1103/PhysRevA.104.032403} {\bibfield  {journal} {\bibinfo  {journal} {Phys. Rev. A}\ }\textbf {\bibinfo {volume} {104}},\ \bibinfo {pages} {032403} (\bibinfo {year} {2021})}\BibitemShut {NoStop}%
\bibitem [{\citenamefont {Nielsen}\ and\ \citenamefont {Chuang}(2010)}]{QIC-book}%
  \BibitemOpen
  \bibfield  {author} {\bibinfo {author} {\bibfnamefont {M.~A.}\ \bibnamefont {Nielsen}}\ and\ \bibinfo {author} {\bibfnamefont {I.~L.}\ \bibnamefont {Chuang}},\ }\href@noop {} {\emph {\bibinfo {title} {Quantum Computation and Quantum Information}}},\ \bibinfo {edition} {10th}\ ed.\ (\bibinfo  {publisher} {Cambridge University Press},\ \bibinfo {year} {2010})\BibitemShut {NoStop}%
\bibitem [{\citenamefont {Ali}\ \emph {et~al.}(2009)\citenamefont {Ali}, \citenamefont {Carmeli}, \citenamefont {Heinosaari},\ and\ \citenamefont {Toigo}}]{Ali2009FP}%
  \BibitemOpen
  \bibfield  {author} {\bibinfo {author} {\bibfnamefont {S.~T.}\ \bibnamefont {Ali}}, \bibinfo {author} {\bibfnamefont {C.}~\bibnamefont {Carmeli}}, \bibinfo {author} {\bibfnamefont {T.}~\bibnamefont {Heinosaari}},\ and\ \bibinfo {author} {\bibfnamefont {A.}~\bibnamefont {Toigo}},\ }\bibfield  {title} {\bibinfo {title} {Commutative povms and fuzzy observables},\ }\href {https://doi.org/10.1007/s10701-009-9292-y} {\bibfield  {journal} {\bibinfo  {journal} {Found. Phys.}\ }\textbf {\bibinfo {volume} {39}},\ \bibinfo {pages} {593} (\bibinfo {year} {2009})}\BibitemShut {NoStop}%
\bibitem [{\citenamefont {Heinosaari}\ \emph {et~al.}(2015)\citenamefont {Heinosaari}, \citenamefont {Kiukas}, \citenamefont {Reitzner},\ and\ \citenamefont {Schultz}}]{Heinosaari2015}%
  \BibitemOpen
  \bibfield  {author} {\bibinfo {author} {\bibfnamefont {T.}~\bibnamefont {Heinosaari}}, \bibinfo {author} {\bibfnamefont {J.}~\bibnamefont {Kiukas}}, \bibinfo {author} {\bibfnamefont {D.}~\bibnamefont {Reitzner}},\ and\ \bibinfo {author} {\bibfnamefont {J.}~\bibnamefont {Schultz}},\ }\bibfield  {title} {\bibinfo {title} {Incompatibility breaking quantum channels},\ }\href {https://doi.org/10.1088/1751-8113/48/43/435301} {\bibfield  {journal} {\bibinfo  {journal} {J. Phys. A: Math. Theor.}\ }\textbf {\bibinfo {volume} {48}},\ \bibinfo {pages} {435301} (\bibinfo {year} {2015})}\BibitemShut {NoStop}%
\bibitem [{\citenamefont {Kiukas}\ \emph {et~al.}(2017)\citenamefont {Kiukas}, \citenamefont {Budroni}, \citenamefont {Uola},\ and\ \citenamefont {Pellonp\"a\"a}}]{Kiukas2017PRA}%
  \BibitemOpen
  \bibfield  {author} {\bibinfo {author} {\bibfnamefont {J.}~\bibnamefont {Kiukas}}, \bibinfo {author} {\bibfnamefont {C.}~\bibnamefont {Budroni}}, \bibinfo {author} {\bibfnamefont {R.}~\bibnamefont {Uola}},\ and\ \bibinfo {author} {\bibfnamefont {J.-P.}\ \bibnamefont {Pellonp\"a\"a}},\ }\bibfield  {title} {\bibinfo {title} {Continuous-variable steering and incompatibility via state-channel duality},\ }\href {https://doi.org/10.1103/PhysRevA.96.042331} {\bibfield  {journal} {\bibinfo  {journal} {Phys. Rev. A}\ }\textbf {\bibinfo {volume} {96}},\ \bibinfo {pages} {042331} (\bibinfo {year} {2017})}\BibitemShut {NoStop}%
\bibitem [{\citenamefont {Ku}\ \emph {et~al.}(2023)\citenamefont {Ku}, \citenamefont {Lee}, \citenamefont {Lai}, \citenamefont {Lin},\ and\ \citenamefont {Chen}}]{Ku2023PRA}%
  \BibitemOpen
  \bibfield  {author} {\bibinfo {author} {\bibfnamefont {H.-Y.}\ \bibnamefont {Ku}}, \bibinfo {author} {\bibfnamefont {K.-Y.}\ \bibnamefont {Lee}}, \bibinfo {author} {\bibfnamefont {P.-R.}\ \bibnamefont {Lai}}, \bibinfo {author} {\bibfnamefont {J.-D.}\ \bibnamefont {Lin}},\ and\ \bibinfo {author} {\bibfnamefont {Y.-N.}\ \bibnamefont {Chen}},\ }\bibfield  {title} {\bibinfo {title} {Coherent activation of a steerability-breaking channel},\ }\href {https://doi.org/10.1103/PhysRevA.107.042415} {\bibfield  {journal} {\bibinfo  {journal} {Phys. Rev. A}\ }\textbf {\bibinfo {volume} {107}},\ \bibinfo {pages} {042415} (\bibinfo {year} {2023})}\BibitemShut {NoStop}%
\bibitem [{\citenamefont {Navascu\'es}\ and\ \citenamefont {Garc\'{\i}a-Pintos}(2015)}]{Navascues2015PRL}%
  \BibitemOpen
  \bibfield  {author} {\bibinfo {author} {\bibfnamefont {M.}~\bibnamefont {Navascu\'es}}\ and\ \bibinfo {author} {\bibfnamefont {L.~P.}\ \bibnamefont {Garc\'{\i}a-Pintos}},\ }\bibfield  {title} {\bibinfo {title} {Nonthermal quantum channels as a thermodynamical resource},\ }\href {https://doi.org/10.1103/PhysRevLett.115.010405} {\bibfield  {journal} {\bibinfo  {journal} {Phys. Rev. Lett.}\ }\textbf {\bibinfo {volume} {115}},\ \bibinfo {pages} {010405} (\bibinfo {year} {2015})}\BibitemShut {NoStop}%
\bibitem [{\citenamefont {Theurer}\ \emph {et~al.}(2019)\citenamefont {Theurer}, \citenamefont {Egloff}, \citenamefont {Zhang},\ and\ \citenamefont {Plenio}}]{Theurer2019PRL}%
  \BibitemOpen
  \bibfield  {author} {\bibinfo {author} {\bibfnamefont {T.}~\bibnamefont {Theurer}}, \bibinfo {author} {\bibfnamefont {D.}~\bibnamefont {Egloff}}, \bibinfo {author} {\bibfnamefont {L.}~\bibnamefont {Zhang}},\ and\ \bibinfo {author} {\bibfnamefont {M.~B.}\ \bibnamefont {Plenio}},\ }\bibfield  {title} {\bibinfo {title} {Quantifying operations with an application to coherence},\ }\href {https://doi.org/10.1103/PhysRevLett.122.190405} {\bibfield  {journal} {\bibinfo  {journal} {Phys. Rev. Lett.}\ }\textbf {\bibinfo {volume} {122}},\ \bibinfo {pages} {190405} (\bibinfo {year} {2019})}\BibitemShut {NoStop}%
\bibitem [{\citenamefont {Liu}\ and\ \citenamefont {Winter}(2019)}]{Liu2019}%
  \BibitemOpen
  \bibfield  {author} {\bibinfo {author} {\bibfnamefont {Z.-W.}\ \bibnamefont {Liu}}\ and\ \bibinfo {author} {\bibfnamefont {A.}~\bibnamefont {Winter}},\ }\href@noop {} {\bibinfo {title} {Resource theories of quantum channels and the universal role of resource erasure}} (\bibinfo {year} {2019}),\ \Eprint {https://arxiv.org/abs/1904.04201} {arXiv:1904.04201 [quant-ph]} \BibitemShut {NoStop}%
\bibitem [{\citenamefont {Liu}\ and\ \citenamefont {Yuan}(2020)}]{Liu2020PRR}%
  \BibitemOpen
  \bibfield  {author} {\bibinfo {author} {\bibfnamefont {Y.}~\bibnamefont {Liu}}\ and\ \bibinfo {author} {\bibfnamefont {X.}~\bibnamefont {Yuan}},\ }\bibfield  {title} {\bibinfo {title} {Operational resource theory of quantum channels},\ }\href {https://doi.org/10.1103/PhysRevResearch.2.012035} {\bibfield  {journal} {\bibinfo  {journal} {Phys. Rev. Res.}\ }\textbf {\bibinfo {volume} {2}},\ \bibinfo {pages} {012035} (\bibinfo {year} {2020})}\BibitemShut {NoStop}%
\bibitem [{\citenamefont {Takagi}\ \emph {et~al.}(2020)\citenamefont {Takagi}, \citenamefont {Wang},\ and\ \citenamefont {Hayashi}}]{Takagi2020PRL}%
  \BibitemOpen
  \bibfield  {author} {\bibinfo {author} {\bibfnamefont {R.}~\bibnamefont {Takagi}}, \bibinfo {author} {\bibfnamefont {K.}~\bibnamefont {Wang}},\ and\ \bibinfo {author} {\bibfnamefont {M.}~\bibnamefont {Hayashi}},\ }\bibfield  {title} {\bibinfo {title} {Application of the resource theory of channels to communication scenarios},\ }\href {https://doi.org/10.1103/PhysRevLett.124.120502} {\bibfield  {journal} {\bibinfo  {journal} {Phys. Rev. Lett.}\ }\textbf {\bibinfo {volume} {124}},\ \bibinfo {pages} {120502} (\bibinfo {year} {2020})}\BibitemShut {NoStop}%
\bibitem [{\citenamefont {Regula}\ and\ \citenamefont {Takagi}(2021{\natexlab{a}})}]{Regula2021PRL}%
  \BibitemOpen
  \bibfield  {author} {\bibinfo {author} {\bibfnamefont {B.}~\bibnamefont {Regula}}\ and\ \bibinfo {author} {\bibfnamefont {R.}~\bibnamefont {Takagi}},\ }\bibfield  {title} {\bibinfo {title} {One-shot manipulation of dynamical quantum resources},\ }\href {https://doi.org/10.1103/PhysRevLett.127.060402} {\bibfield  {journal} {\bibinfo  {journal} {Phys. Rev. Lett.}\ }\textbf {\bibinfo {volume} {127}},\ \bibinfo {pages} {060402} (\bibinfo {year} {2021}{\natexlab{a}})}\BibitemShut {NoStop}%
\bibitem [{\citenamefont {Regula}\ and\ \citenamefont {Takagi}(2021{\natexlab{b}})}]{Regula2021NC}%
  \BibitemOpen
  \bibfield  {author} {\bibinfo {author} {\bibfnamefont {B.}~\bibnamefont {Regula}}\ and\ \bibinfo {author} {\bibfnamefont {R.}~\bibnamefont {Takagi}},\ }\bibfield  {title} {\bibinfo {title} {Fundamental limitations on distillation of quantum channel resources},\ }\href {https://doi.org/10.1038/s41467-021-24699-0} {\bibfield  {journal} {\bibinfo  {journal} {Nat. Commun.}\ }\textbf {\bibinfo {volume} {12}},\ \bibinfo {pages} {4411} (\bibinfo {year} {2021}{\natexlab{b}})}\BibitemShut {NoStop}%
\bibitem [{\citenamefont {Hsieh}(2022)}]{Hsieh-thesis}%
  \BibitemOpen
  \bibfield  {author} {\bibinfo {author} {\bibfnamefont {C.-Y.}\ \bibnamefont {Hsieh}},\ }\emph {\bibinfo {title} {Resource theories of quantum dynamics}},\ \href {https://doi.org/10.5821/dissertation-2117-379454} {Ph.D. thesis},\ \bibinfo  {school} {UPC, Institut de Ci\'encies Fot\'oniques} (\bibinfo {year} {2022})\BibitemShut {NoStop}%
\bibitem [{\citenamefont {Davies}\ and\ \citenamefont {Lewis}(1970)}]{Davies1970CMP}%
  \BibitemOpen
  \bibfield  {author} {\bibinfo {author} {\bibfnamefont {E.~B.}\ \bibnamefont {Davies}}\ and\ \bibinfo {author} {\bibfnamefont {J.~T.}\ \bibnamefont {Lewis}},\ }\bibfield  {title} {\bibinfo {title} {An operational approach to quantum probability},\ }\href {https://doi.org/10.1007/BF01647093} {\bibfield  {journal} {\bibinfo  {journal} {Commun. Math. Phys.}\ }\textbf {\bibinfo {volume} {17}},\ \bibinfo {pages} {239} (\bibinfo {year} {1970})}\BibitemShut {NoStop}%
\bibitem [{\citenamefont {Chitambar}\ and\ \citenamefont {Gour}(2019)}]{ChitambarRMP2019}%
  \BibitemOpen
  \bibfield  {author} {\bibinfo {author} {\bibfnamefont {E.}~\bibnamefont {Chitambar}}\ and\ \bibinfo {author} {\bibfnamefont {G.}~\bibnamefont {Gour}},\ }\bibfield  {title} {\bibinfo {title} {Quantum resource theories},\ }\href {https://doi.org/10.1103/RevModPhys.91.025001} {\bibfield  {journal} {\bibinfo  {journal} {Rev. Mod. Phys.}\ }\textbf {\bibinfo {volume} {91}},\ \bibinfo {pages} {025001} (\bibinfo {year} {2019})}\BibitemShut {NoStop}%
\bibitem [{\citenamefont {Horodecki}\ \emph {et~al.}(2003)\citenamefont {Horodecki}, \citenamefont {Shor},\ and\ \citenamefont {Ruskai}}]{Horodecki2003RevMP}%
  \BibitemOpen
  \bibfield  {author} {\bibinfo {author} {\bibfnamefont {M.}~\bibnamefont {Horodecki}}, \bibinfo {author} {\bibfnamefont {P.~W.}\ \bibnamefont {Shor}},\ and\ \bibinfo {author} {\bibfnamefont {M.~B.}\ \bibnamefont {Ruskai}},\ }\bibfield  {title} {\bibinfo {title} {Entanglement breaking channels},\ }\href {https://doi.org/10.1142/S0129055X03001709} {\bibfield  {journal} {\bibinfo  {journal} {Rev. Math. Phys.}\ }\textbf {\bibinfo {volume} {15}},\ \bibinfo {pages} {629} (\bibinfo {year} {2003})}\BibitemShut {NoStop}%
\bibitem [{\citenamefont {Tabia}\ and\ \citenamefont {Hsieh}(2024)}]{Tabia2024}%
  \BibitemOpen
  \bibfield  {author} {\bibinfo {author} {\bibfnamefont {G.~N.~M.}\ \bibnamefont {Tabia}}\ and\ \bibinfo {author} {\bibfnamefont {C.-Y.}\ \bibnamefont {Hsieh}},\ }\href@noop {} {\bibinfo {title} {Super-activating quantum memory by entanglement-breaking channels}} (\bibinfo {year} {2024}),\ \Eprint {https://arxiv.org/abs/2410.13499} {arXiv:2410.13499 [quant-ph]} \BibitemShut {NoStop}%
\bibitem [{\citenamefont {Hsieh}\ \emph {et~al.}(2016)\citenamefont {Hsieh}, \citenamefont {Liang},\ and\ \citenamefont {Lee}}]{Hsieh2016PRA}%
  \BibitemOpen
  \bibfield  {author} {\bibinfo {author} {\bibfnamefont {C.-Y.}\ \bibnamefont {Hsieh}}, \bibinfo {author} {\bibfnamefont {Y.-C.}\ \bibnamefont {Liang}},\ and\ \bibinfo {author} {\bibfnamefont {R.-K.}\ \bibnamefont {Lee}},\ }\bibfield  {title} {\bibinfo {title} {Quantum steerability: Characterization, quantification, superactivation, and unbounded amplification},\ }\href {https://doi.org/10.1103/PhysRevA.94.062120} {\bibfield  {journal} {\bibinfo  {journal} {Phys. Rev. A}\ }\textbf {\bibinfo {volume} {94}},\ \bibinfo {pages} {062120} (\bibinfo {year} {2016})}\BibitemShut {NoStop}%
\bibitem [{\citenamefont {Buscemi}\ and\ \citenamefont {Datta}(2010)}]{Buscemi2010ITIT}%
  \BibitemOpen
  \bibfield  {author} {\bibinfo {author} {\bibfnamefont {F.}~\bibnamefont {Buscemi}}\ and\ \bibinfo {author} {\bibfnamefont {N.}~\bibnamefont {Datta}},\ }\bibfield  {title} {\bibinfo {title} {The quantum capacity of channels with arbitrarily correlated noise},\ }\href@noop {} {\bibfield  {journal} {\bibinfo  {journal} {IEEE Trans. Inf. Theory}\ }\textbf {\bibinfo {volume} {56}},\ \bibinfo {pages} {1447} (\bibinfo {year} {2010})}\BibitemShut {NoStop}%
\bibitem [{\citenamefont {Horodecki}\ \emph {et~al.}(1999)\citenamefont {Horodecki}, \citenamefont {Horodecki},\ and\ \citenamefont {Horodecki}}]{Horodecki1999PRA}%
  \BibitemOpen
  \bibfield  {author} {\bibinfo {author} {\bibfnamefont {M.}~\bibnamefont {Horodecki}}, \bibinfo {author} {\bibfnamefont {P.}~\bibnamefont {Horodecki}},\ and\ \bibinfo {author} {\bibfnamefont {R.}~\bibnamefont {Horodecki}},\ }\bibfield  {title} {\bibinfo {title} {General teleportation channel, singlet fraction, and quasidistillation},\ }\href {https://doi.org/10.1103/PhysRevA.60.1888} {\bibfield  {journal} {\bibinfo  {journal} {Phys. Rev. A}\ }\textbf {\bibinfo {volume} {60}},\ \bibinfo {pages} {1888} (\bibinfo {year} {1999})}\BibitemShut {NoStop}%
\bibitem [{\citenamefont {Horodecki}\ \emph {et~al.}(2009)\citenamefont {Horodecki}, \citenamefont {Horodecki}, \citenamefont {Horodecki},\ and\ \citenamefont {Horodecki}}]{Horodecki2009RMP}%
  \BibitemOpen
  \bibfield  {author} {\bibinfo {author} {\bibfnamefont {R.}~\bibnamefont {Horodecki}}, \bibinfo {author} {\bibfnamefont {P.}~\bibnamefont {Horodecki}}, \bibinfo {author} {\bibfnamefont {M.}~\bibnamefont {Horodecki}},\ and\ \bibinfo {author} {\bibfnamefont {K.}~\bibnamefont {Horodecki}},\ }\bibfield  {title} {\bibinfo {title} {Quantum entanglement},\ }\href {https://doi.org/10.1103/RevModPhys.81.865} {\bibfield  {journal} {\bibinfo  {journal} {Rev. Mod. Phys.}\ }\textbf {\bibinfo {volume} {81}},\ \bibinfo {pages} {865} (\bibinfo {year} {2009})}\BibitemShut {NoStop}%
\bibitem [{\citenamefont {Renner}(2024)}]{RennerPRL2024}%
  \BibitemOpen
  \bibfield  {author} {\bibinfo {author} {\bibfnamefont {M.~J.}\ \bibnamefont {Renner}},\ }\bibfield  {title} {\bibinfo {title} {Compatibility of generalized noisy qubit measurements},\ }\href {https://doi.org/10.1103/PhysRevLett.132.250202} {\bibfield  {journal} {\bibinfo  {journal} {Phys. Rev. Lett.}\ }\textbf {\bibinfo {volume} {132}},\ \bibinfo {pages} {250202} (\bibinfo {year} {2024})}\BibitemShut {NoStop}%
\bibitem [{\citenamefont {Zhang}\ and\ \citenamefont {Chitambar}(2024)}]{ZhangPRL2024}%
  \BibitemOpen
  \bibfield  {author} {\bibinfo {author} {\bibfnamefont {Y.}~\bibnamefont {Zhang}}\ and\ \bibinfo {author} {\bibfnamefont {E.}~\bibnamefont {Chitambar}},\ }\bibfield  {title} {\bibinfo {title} {Exact steering bound for two-qubit werner states},\ }\href {https://doi.org/10.1103/PhysRevLett.132.250201} {\bibfield  {journal} {\bibinfo  {journal} {Phys. Rev. Lett.}\ }\textbf {\bibinfo {volume} {132}},\ \bibinfo {pages} {250201} (\bibinfo {year} {2024})}\BibitemShut {NoStop}%
\bibitem [{\citenamefont {Palazuelos}(2012)}]{Palazuelos2012PRL}%
  \BibitemOpen
  \bibfield  {author} {\bibinfo {author} {\bibfnamefont {C.}~\bibnamefont {Palazuelos}},\ }\bibfield  {title} {\bibinfo {title} {Superactivation of quantum nonlocality},\ }\href {https://doi.org/10.1103/PhysRevLett.109.190401} {\bibfield  {journal} {\bibinfo  {journal} {Phys. Rev. Lett.}\ }\textbf {\bibinfo {volume} {109}},\ \bibinfo {pages} {190401} (\bibinfo {year} {2012})}\BibitemShut {NoStop}%
\bibitem [{\citenamefont {Li}\ \emph {et~al.}(2021)\citenamefont {Li}, \citenamefont {Fang}, \citenamefont {Zhang}, \citenamefont {Tabia}, \citenamefont {Lu},\ and\ \citenamefont {Liang}}]{Li2021PRR}%
  \BibitemOpen
  \bibfield  {author} {\bibinfo {author} {\bibfnamefont {J.-Y.}\ \bibnamefont {Li}}, \bibinfo {author} {\bibfnamefont {X.-X.}\ \bibnamefont {Fang}}, \bibinfo {author} {\bibfnamefont {T.}~\bibnamefont {Zhang}}, \bibinfo {author} {\bibfnamefont {G.~N.~M.}\ \bibnamefont {Tabia}}, \bibinfo {author} {\bibfnamefont {H.}~\bibnamefont {Lu}},\ and\ \bibinfo {author} {\bibfnamefont {Y.-C.}\ \bibnamefont {Liang}},\ }\bibfield  {title} {\bibinfo {title} {Activating hidden teleportation power: Theory and experiment},\ }\href {https://doi.org/10.1103/PhysRevResearch.3.023045} {\bibfield  {journal} {\bibinfo  {journal} {Phys. Rev. Res.}\ }\textbf {\bibinfo {volume} {3}},\ \bibinfo {pages} {023045} (\bibinfo {year} {2021})}\BibitemShut {NoStop}%
\bibitem [{\citenamefont {Cavalcanti}\ \emph {et~al.}(2013)\citenamefont {Cavalcanti}, \citenamefont {Ac\'{\i}n}, \citenamefont {Brunner},\ and\ \citenamefont {V\'ertesi}}]{CavalcantiPRA2013}%
  \BibitemOpen
  \bibfield  {author} {\bibinfo {author} {\bibfnamefont {D.}~\bibnamefont {Cavalcanti}}, \bibinfo {author} {\bibfnamefont {A.}~\bibnamefont {Ac\'{\i}n}}, \bibinfo {author} {\bibfnamefont {N.}~\bibnamefont {Brunner}},\ and\ \bibinfo {author} {\bibfnamefont {T.}~\bibnamefont {V\'ertesi}},\ }\bibfield  {title} {\bibinfo {title} {All quantum states useful for teleportation are nonlocal resources},\ }\href {https://doi.org/10.1103/PhysRevA.87.042104} {\bibfield  {journal} {\bibinfo  {journal} {Phys. Rev. A}\ }\textbf {\bibinfo {volume} {87}},\ \bibinfo {pages} {042104} (\bibinfo {year} {2013})}\BibitemShut {NoStop}%
\bibitem [{\citenamefont {Wiseman}\ \emph {et~al.}(2007)\citenamefont {Wiseman}, \citenamefont {Jones},\ and\ \citenamefont {Doherty}}]{Wiseman2007PRL}%
  \BibitemOpen
  \bibfield  {author} {\bibinfo {author} {\bibfnamefont {H.~M.}\ \bibnamefont {Wiseman}}, \bibinfo {author} {\bibfnamefont {S.~J.}\ \bibnamefont {Jones}},\ and\ \bibinfo {author} {\bibfnamefont {A.~C.}\ \bibnamefont {Doherty}},\ }\bibfield  {title} {\bibinfo {title} {Steering, entanglement, nonlocality, and the {E}instein-{P}odolsky-{R}osen paradox},\ }\href {https://doi.org/10.1103/PhysRevLett.98.140402} {\bibfield  {journal} {\bibinfo  {journal} {Phys. Rev. Lett.}\ }\textbf {\bibinfo {volume} {98}},\ \bibinfo {pages} {140402} (\bibinfo {year} {2007})}\BibitemShut {NoStop}%
\bibitem [{\citenamefont {Ducuara}\ and\ \citenamefont {Skrzypczyk}(2020)}]{Ducuara2020PRL}%
  \BibitemOpen
  \bibfield  {author} {\bibinfo {author} {\bibfnamefont {A.~F.}\ \bibnamefont {Ducuara}}\ and\ \bibinfo {author} {\bibfnamefont {P.}~\bibnamefont {Skrzypczyk}},\ }\bibfield  {title} {\bibinfo {title} {Operational interpretation of weight-based resource quantifiers in convex quantum resource theories},\ }\href {https://doi.org/10.1103/PhysRevLett.125.110401} {\bibfield  {journal} {\bibinfo  {journal} {Phys. Rev. Lett.}\ }\textbf {\bibinfo {volume} {125}},\ \bibinfo {pages} {110401} (\bibinfo {year} {2020})}\BibitemShut {NoStop}%
\bibitem [{\citenamefont {Ducuara}\ and\ \citenamefont {Skrzypczyk}(2022)}]{Ducuara2022PRXQ}%
  \BibitemOpen
  \bibfield  {author} {\bibinfo {author} {\bibfnamefont {A.~F.}\ \bibnamefont {Ducuara}}\ and\ \bibinfo {author} {\bibfnamefont {P.}~\bibnamefont {Skrzypczyk}},\ }\bibfield  {title} {\bibinfo {title} {Characterization of quantum betting tasks in terms of arimoto mutual information},\ }\href {https://doi.org/10.1103/PRXQuantum.3.020366} {\bibfield  {journal} {\bibinfo  {journal} {PRX Quantum}\ }\textbf {\bibinfo {volume} {3}},\ \bibinfo {pages} {020366} (\bibinfo {year} {2022})}\BibitemShut {NoStop}%
\bibitem [{\citenamefont {Ducuara}\ \emph {et~al.}(2023)\citenamefont {Ducuara}, \citenamefont {Skrzypczyk}, \citenamefont {Buscemi}, \citenamefont {Sidajaya},\ and\ \citenamefont {Scarani}}]{Ducuara2023PRL}%
  \BibitemOpen
  \bibfield  {author} {\bibinfo {author} {\bibfnamefont {A.~F.}\ \bibnamefont {Ducuara}}, \bibinfo {author} {\bibfnamefont {P.}~\bibnamefont {Skrzypczyk}}, \bibinfo {author} {\bibfnamefont {F.}~\bibnamefont {Buscemi}}, \bibinfo {author} {\bibfnamefont {P.}~\bibnamefont {Sidajaya}},\ and\ \bibinfo {author} {\bibfnamefont {V.}~\bibnamefont {Scarani}},\ }\bibfield  {title} {\bibinfo {title} {Maxwell's demon walks into wall street: Stochastic thermodynamics meets expected utility theory},\ }\href {https://doi.org/10.1103/PhysRevLett.131.197103} {\bibfield  {journal} {\bibinfo  {journal} {Phys. Rev. Lett.}\ }\textbf {\bibinfo {volume} {131}},\ \bibinfo {pages} {197103} (\bibinfo {year} {2023})}\BibitemShut {NoStop}%
\bibitem [{\citenamefont {Stratton}\ \emph {et~al.}(2024{\natexlab{b}})\citenamefont {Stratton}, \citenamefont {Hsieh},\ and\ \citenamefont {Skrzypczyk}}]{stratton2024}%
  \BibitemOpen
  \bibfield  {author} {\bibinfo {author} {\bibfnamefont {B.}~\bibnamefont {Stratton}}, \bibinfo {author} {\bibfnamefont {C.-Y.}\ \bibnamefont {Hsieh}},\ and\ \bibinfo {author} {\bibfnamefont {P.}~\bibnamefont {Skrzypczyk}},\ }\bibfield  {title} {\bibinfo {title} {Operational interpretation of the choi rank through exclusion tasks},\ }\href {https://doi.org/10.1103/PhysRevA.110.L050601} {\bibfield  {journal} {\bibinfo  {journal} {Phys. Rev. A}\ }\textbf {\bibinfo {volume} {110}},\ \bibinfo {pages} {L050601} (\bibinfo {year} {2024}{\natexlab{b}})}\BibitemShut {NoStop}%
\bibitem [{\citenamefont {Mitra}\ and\ \citenamefont {Farkas}(2022)}]{Mitra2022PRA}%
  \BibitemOpen
  \bibfield  {author} {\bibinfo {author} {\bibfnamefont {A.}~\bibnamefont {Mitra}}\ and\ \bibinfo {author} {\bibfnamefont {M.}~\bibnamefont {Farkas}},\ }\bibfield  {title} {\bibinfo {title} {Compatibility of quantum instruments},\ }\href {https://doi.org/10.1103/PhysRevA.105.052202} {\bibfield  {journal} {\bibinfo  {journal} {Phys. Rev. A}\ }\textbf {\bibinfo {volume} {105}},\ \bibinfo {pages} {052202} (\bibinfo {year} {2022})}\BibitemShut {NoStop}%
\bibitem [{\citenamefont {Mitra}\ and\ \citenamefont {Farkas}(2023)}]{Mitra2023PRA}%
  \BibitemOpen
  \bibfield  {author} {\bibinfo {author} {\bibfnamefont {A.}~\bibnamefont {Mitra}}\ and\ \bibinfo {author} {\bibfnamefont {M.}~\bibnamefont {Farkas}},\ }\bibfield  {title} {\bibinfo {title} {Characterizing and quantifying the incompatibility of quantum instruments},\ }\href {https://doi.org/10.1103/PhysRevA.107.032217} {\bibfield  {journal} {\bibinfo  {journal} {Phys. Rev. A}\ }\textbf {\bibinfo {volume} {107}},\ \bibinfo {pages} {032217} (\bibinfo {year} {2023})}\BibitemShut {NoStop}%
\bibitem [{\citenamefont {Lepp{\"{a}}j{\"{a}}rvi}\ and\ \citenamefont {Sedl{\'{a}}k}(2024)}]{Leppajarvi2024incompatibilityof}%
  \BibitemOpen
  \bibfield  {author} {\bibinfo {author} {\bibfnamefont {L.}~\bibnamefont {Lepp{\"{a}}j{\"{a}}rvi}}\ and\ \bibinfo {author} {\bibfnamefont {M.}~\bibnamefont {Sedl{\'{a}}k}},\ }\bibfield  {title} {\bibinfo {title} {Incompatibility of quantum instruments},\ }\href {https://doi.org/10.22331/q-2024-02-12-1246} {\bibfield  {journal} {\bibinfo  {journal} {{Quantum}}\ }\textbf {\bibinfo {volume} {8}},\ \bibinfo {pages} {1246} (\bibinfo {year} {2024})}\BibitemShut {NoStop}%
\bibitem [{\citenamefont {Haapasalo}\ \emph {et~al.}(2021)\citenamefont {Haapasalo}, \citenamefont {Kraft}, \citenamefont {Miklin},\ and\ \citenamefont {Uola}}]{Haapasalo2021}%
  \BibitemOpen
  \bibfield  {author} {\bibinfo {author} {\bibfnamefont {E.}~\bibnamefont {Haapasalo}}, \bibinfo {author} {\bibfnamefont {T.}~\bibnamefont {Kraft}}, \bibinfo {author} {\bibfnamefont {N.}~\bibnamefont {Miklin}},\ and\ \bibinfo {author} {\bibfnamefont {R.}~\bibnamefont {Uola}},\ }\bibfield  {title} {\bibinfo {title} {Quantum marginal problem and incompatibility},\ }\href {https://doi.org/10.22331/q-2021-06-15-476} {\bibfield  {journal} {\bibinfo  {journal} {{Quantum}}\ }\textbf {\bibinfo {volume} {5}},\ \bibinfo {pages} {476} (\bibinfo {year} {2021})}\BibitemShut {NoStop}%
\bibitem [{\citenamefont {Hsieh}\ \emph {et~al.}(2022)\citenamefont {Hsieh}, \citenamefont {Lostaglio},\ and\ \citenamefont {Ac\'{\i}n}}]{Hsieh2022PRR}%
  \BibitemOpen
  \bibfield  {author} {\bibinfo {author} {\bibfnamefont {C.-Y.}\ \bibnamefont {Hsieh}}, \bibinfo {author} {\bibfnamefont {M.}~\bibnamefont {Lostaglio}},\ and\ \bibinfo {author} {\bibfnamefont {A.}~\bibnamefont {Ac\'{\i}n}},\ }\bibfield  {title} {\bibinfo {title} {Quantum channel marginal problem},\ }\href {https://doi.org/10.1103/PhysRevResearch.4.013249} {\bibfield  {journal} {\bibinfo  {journal} {Phys. Rev. Res.}\ }\textbf {\bibinfo {volume} {4}},\ \bibinfo {pages} {013249} (\bibinfo {year} {2022})}\BibitemShut {NoStop}%
\bibitem [{\citenamefont {Hsieh}\ \emph {et~al.}(2024)\citenamefont {Hsieh}, \citenamefont {Tabia}, \citenamefont {Yin},\ and\ \citenamefont {Liang}}]{Hsieh2024Quantum}%
  \BibitemOpen
  \bibfield  {author} {\bibinfo {author} {\bibfnamefont {C.-Y.}\ \bibnamefont {Hsieh}}, \bibinfo {author} {\bibfnamefont {G.~N.~M.}\ \bibnamefont {Tabia}}, \bibinfo {author} {\bibfnamefont {Y.-C.}\ \bibnamefont {Yin}},\ and\ \bibinfo {author} {\bibfnamefont {Y.-C.}\ \bibnamefont {Liang}},\ }\bibfield  {title} {\bibinfo {title} {Resource marginal problems},\ }\href {https://doi.org/10.22331/q-2024-05-22-1353} {\bibfield  {journal} {\bibinfo  {journal} {{Quantum}}\ }\textbf {\bibinfo {volume} {8}},\ \bibinfo {pages} {1353} (\bibinfo {year} {2024})}\BibitemShut {NoStop}%
\bibitem [{\citenamefont {Tabia}\ \emph {et~al.}(2022)\citenamefont {Tabia}, \citenamefont {Chen}, \citenamefont {Hsieh}, \citenamefont {Yin},\ and\ \citenamefont {Liang}}]{Tabia2022npjQI}%
  \BibitemOpen
  \bibfield  {author} {\bibinfo {author} {\bibfnamefont {G.~N.~M.}\ \bibnamefont {Tabia}}, \bibinfo {author} {\bibfnamefont {K.-S.}\ \bibnamefont {Chen}}, \bibinfo {author} {\bibfnamefont {C.-Y.}\ \bibnamefont {Hsieh}}, \bibinfo {author} {\bibfnamefont {Y.-C.}\ \bibnamefont {Yin}},\ and\ \bibinfo {author} {\bibfnamefont {Y.-C.}\ \bibnamefont {Liang}},\ }\bibfield  {title} {\bibinfo {title} {Entanglement transitivity problems},\ }\href {https://doi.org/10.1038/s41534-022-00616-1} {\bibfield  {journal} {\bibinfo  {journal} {npj Quantum Inf.}\ }\textbf {\bibinfo {volume} {8}},\ \bibinfo {pages} {98} (\bibinfo {year} {2022})}\BibitemShut {NoStop}%
\bibitem [{\citenamefont {Aharonov}\ \emph {et~al.}(1998)\citenamefont {Aharonov}, \citenamefont {Kitaev},\ and\ \citenamefont {Nisan}}]{Aharonov1998}%
  \BibitemOpen
  \bibfield  {author} {\bibinfo {author} {\bibfnamefont {D.}~\bibnamefont {Aharonov}}, \bibinfo {author} {\bibfnamefont {A.}~\bibnamefont {Kitaev}},\ and\ \bibinfo {author} {\bibfnamefont {N.}~\bibnamefont {Nisan}},\ }\href@noop {} {\bibinfo {title} {Quantum circuits with mixed states, in {P}roceedings of the {T}hirtieth {A}nnual {ACM} {S}ymposium on {T}heory of {C}omputing, {STOC} ’98 ({A}ssociation for {C}omputing {M}achinery, {N}ew {Y}ork, {NY}, {USA}, 1998), p. 20.}} (\bibinfo {year} {1998}),\ \Eprint {https://arxiv.org/abs/quant-ph/9806029} {arXiv:quant-ph/9806029 [quant-ph]} \BibitemShut {NoStop}%
\bibitem [{\citenamefont {Watrous}(2018)}]{watrous_2018}%
  \BibitemOpen
  \bibfield  {author} {\bibinfo {author} {\bibfnamefont {J.}~\bibnamefont {Watrous}},\ }\href {https://doi.org/10.1017/9781316848142} {\emph {\bibinfo {title} {The Theory of Quantum Information}}}\ (\bibinfo  {publisher} {Cambridge University Press},\ \bibinfo {year} {2018})\BibitemShut {NoStop}%
\bibitem [{\citenamefont {Takagi}\ and\ \citenamefont {Regula}(2019)}]{Takagi2019PRX}%
  \BibitemOpen
  \bibfield  {author} {\bibinfo {author} {\bibfnamefont {R.}~\bibnamefont {Takagi}}\ and\ \bibinfo {author} {\bibfnamefont {B.}~\bibnamefont {Regula}},\ }\bibfield  {title} {\bibinfo {title} {General resource theories in quantum mechanics and beyond: Operational characterization via discrimination tasks},\ }\href {https://doi.org/10.1103/PhysRevX.9.031053} {\bibfield  {journal} {\bibinfo  {journal} {Phys. Rev. X}\ }\textbf {\bibinfo {volume} {9}},\ \bibinfo {pages} {031053} (\bibinfo {year} {2019})}\BibitemShut {NoStop}%
\bibitem [{\citenamefont {Sion}(1958)}]{Sion1958}%
  \BibitemOpen
  \bibfield  {author} {\bibinfo {author} {\bibfnamefont {M.}~\bibnamefont {Sion}},\ }\bibfield  {title} {\bibinfo {title} {On general minimax theorems},\ }\href@noop {} {\bibfield  {journal} {\bibinfo  {journal} {Pac. J. Math.}\ }\textbf {\bibinfo {volume} {8}},\ \bibinfo {pages} {171} (\bibinfo {year} {1958})}\BibitemShut {NoStop}%
\bibitem [{\citenamefont {Komiya}(1988)}]{Hidetoshi1988}%
  \BibitemOpen
  \bibfield  {author} {\bibinfo {author} {\bibfnamefont {H.}~\bibnamefont {Komiya}},\ }\bibfield  {title} {\bibinfo {title} {{Elementary proof for Sion's minimax theorem}},\ }\href {https://doi.org/10.2996/kmj/1138038812} {\bibfield  {journal} {\bibinfo  {journal} {Kodai Math. J.}\ }\textbf {\bibinfo {volume} {11}},\ \bibinfo {pages} {5} (\bibinfo {year} {1988})}\BibitemShut {NoStop}%
\bibitem [{\citenamefont {Choi}(1975)}]{Choi1975}%
  \BibitemOpen
  \bibfield  {author} {\bibinfo {author} {\bibfnamefont {M.-D.}\ \bibnamefont {Choi}},\ }\bibfield  {title} {\bibinfo {title} {Completely positive linear maps on complex matrices},\ }\href {https://doi.org/https://doi.org/10.1016/0024-3795(75)90075-0} {\bibfield  {journal} {\bibinfo  {journal} {Linear Algebr. Appl.}\ }\textbf {\bibinfo {volume} {10}},\ \bibinfo {pages} {285} (\bibinfo {year} {1975})}\BibitemShut {NoStop}%
\bibitem [{\citenamefont {Jamio{\l}kowski}(1972)}]{Jamiolkowski1972}%
  \BibitemOpen
  \bibfield  {author} {\bibinfo {author} {\bibfnamefont {A.}~\bibnamefont {Jamio{\l}kowski}},\ }\bibfield  {title} {\bibinfo {title} {Linear transformations which preserve trace and positive semidefiniteness of operators},\ }\href {https://doi.org/https://doi.org/10.1016/0034-4877(72)90011-0} {\bibfield  {journal} {\bibinfo  {journal} {Rep. Math. Phys.}\ }\textbf {\bibinfo {volume} {3}},\ \bibinfo {pages} {275} (\bibinfo {year} {1972})}\BibitemShut {NoStop}%
\end{thebibliography}%

\end{document}